# Designing Perovskites for Chemical Looping Hydrogen Production

Ivana Staničić[1,2*], Alfred Juhlin Onbeck[2], Anders Hellman[3], Tobias Mattisson[2], Magnus Rydén[1,2]

[1] Wallenberg Initiative Materials Science for Sustainability, Department of Environmental and Energy Sciences, Chalmers University of Technology, Chalmersplatsen 4, Gothenburg, SE-41296, Sweden
[2] Department of Environmental and Energy Sciences, Chalmers University of Technology, Chalmersplatsen 4, Gothenburg, SE-41296, Sweden
[3] Department of Physics and Astronomy, Chalmers University of Technology, Chalmersplatsen 4, Gothenburg, SE-41296, Sweden
*Corresponding author, stanicic@chalmers.se

# Abstract

Perovskites are promising materials for chemical looping hydrogen production (CLHP) due to their compositional flexibility and tunable redox properties. However, cation substitution opens a vast compositional space, making screening for new high-performing materials computationally demanding. Here, an autonomous active-learning framework combining a Wasserstein autoencoder (WAE) with a deep ensemble surrogate model was used to search for perovskites with a high hydrogen production capacity (HPC). In seven independent runs of 50 cycles, the mean HPC of the acquired compositions rose from 0.41 mmol $H_2$ $g^{-1}$ in the starting set to 2.05 mmol $H_2$ $g^{-1}$, and the best compositions reached 2.65–2.99 mmol $H_2$ $g^{-1}$. All runs enriched Sn on the B-site, mostly combined with Ti and with Mg, Ce, K and Y on the A-site. Twenty diverse candidates (2.22–2.84 mmol $H_2$ $g^{-1}$) were then selected by a multi-objective ranking that also accounts for raw-material price, supply resilience and potential environmental and health impacts. The approach links materials discovery with practical deployment constraints with a methodology transferable to other chemical looping processes.

# 1. Introduction

Hydrogen is widely recognized as a key energy carrier for decarbonizing industry, power generation, and transport. Achieving a scalable hydrogen economy, however, depends on the development of production pathways that combine low greenhouse-gas emissions with compatibility to large-scale, continuous operation. [1] Electrolysis is expected to play an important role in future low-carbon hydrogen production, particularly through the integration of renewable and other low-carbon electricity sources. [2] Thermochemical production pathways integrated with concentrated heat sources and carbon capture have the potential to achieve net-zero or even negative emissions, making them an important complement to electrolysis in future hydrogen production systems. [3]

Chemical looping water splitting (CLWS), also known as chemical looping hydrogen production (CLHP), is a promising thermochemical route to continuous hydrogen production through cyclic redox reactions of a solid oxygen carrier (OC) using interconnected fluidized bed reactors as illustrated in Figure 1. By spatially separating reduction and oxidation steps, CLHP intrinsically avoids direct mixing of fuel and oxidant, eliminates nitrogen dilution, and simplifies downstream gas separation.

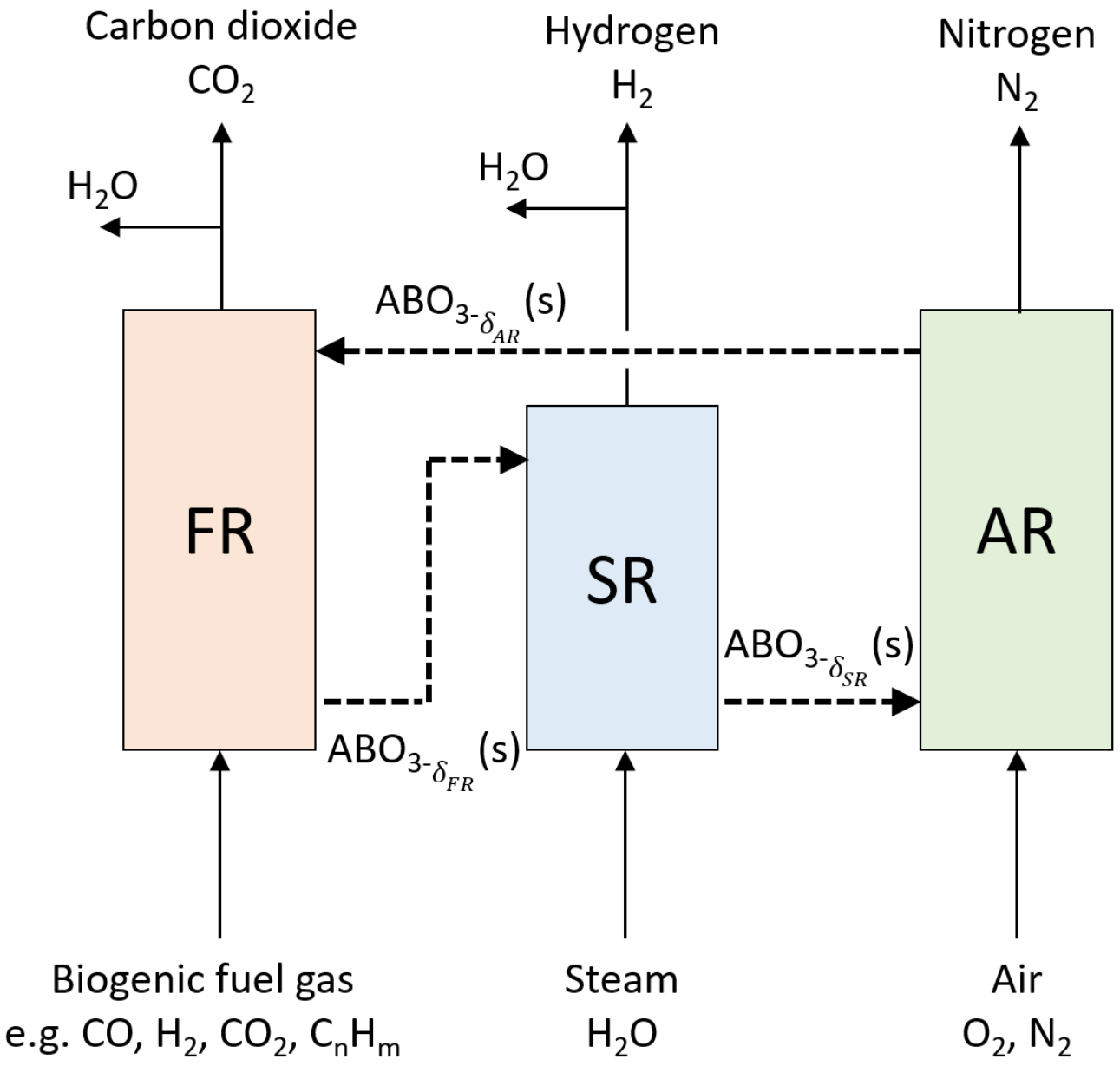

*Figure 1 Schematic illustration of Chemical Looping Hydrogen Production (CLHP) for continuous production of hydrogen using a perovskite ($ABO_{3-\delta}$) as bed material. The fuel is envisioned to be a low-grade gaseous fuel of biogenic origin. The process consists of three interconnected fluidized beds i.e. fuel reactor (FR), steam reactor (SR), and air reactor (AR).*

In CLHP, the oxygen carrier must satisfy three sequential thermodynamic requirements (Figure 1). In the FR, it must release lattice oxygen to the fuel, forming oxygen vacancies ($\delta_{FR}$). In the SR, the reduced carrier is re-oxidized by steam, producing $H_2$. Because steam is a weaker oxidant than air, some of the vacancies are refilled ($\delta_{SR}$). Finally, the material is fully re-oxidized in the AR, which closes the loop and supplies heat to the process. The hydrogen produced per cycle is therefore governed by the difference in oxygen non-stoichiometry between the FR and SR, $\Delta\delta = \delta_{FR} - \delta_{SR}$. This requires vacancy formation free energies that are low enough for reduction in the FR, yet high enough for re-oxidation by steam in the SR. In addition, the carrier must retain its perovskite structure under strongly reducing conditions rather than decomposing into, e.g., brownmillerite or spinel phases.

## 1.1. Oxygen Carriers for Chemical Looping Hydrogen Production

In thermochemical water splitting, hydrogen is produced in a cycle of at least two steps using a redox-active metal oxide, the oxygen carrier (OC). The OC is reduced in the first step and re-oxidized by steam in the second, so that hydrogen is obtained separately from the reduction products. Two process variants are commonly distinguished. In thermal (solar thermochemical) water splitting, the oxide is reduced at high temperature, typically 1300–1500 °C, under inert gas or low oxygen partial pressure, and re-oxidized by steam at an equal or lower temperature. [4] In chemical looping, the oxide is instead reduced by a fuel such as $CH_4$, CO, $H_2$ or biogas, which lowers the operating temperature to about 800–1000 °C. [5] Perovskite oxides ($ABO_{3-\delta}$) are attractive oxygen carriers for both variants because their oxygen non-stoichiometry, redox thermodynamics and kinetics can be tuned over a wide range by substitution on the A and B sites while the crystal structure is retained.

The efficiency, durability, and economic viability of CLHP are largely influenced by the oxygen carrier material. [6] Conventional OCs, such as iron oxides, sinter during deep reduction causing loss of reactivity under cyclic operation, constraining both hydrogen yield and long-term performance. [7] Perovskite materials ($ABO_{3-\delta}$) offer a promising solution to overcome these limitations. Their flexible crystal structure allows extensive substitution on both cation sublattices, enabling tuning of redox thermodynamics, oxygen non-stoichiometry (δ), and phase stability. [3] Consequently, perovskites have attracted considerable attention as oxygen carriers for CLHP, and high-throughput computational screening has successfully identified numerous promising compositions. [8] However, their compositional flexibility also creates a vast combinatorial design space comprising millions of chemically plausible compositions. The challenge is, therefore, to identify compositions that combine phase stability and hydrogen production capacities and can be synthesized at scale. As the compositional complexity increases, high-throughput screening quickly becomes computationally expensive, which motivates the development of more efficient materials discovery strategies. [9, 10]

### 1.1.1. Thermal reduction

Ceria is the benchmark material for thermal water splitting. One perovskite reported to outperform it was Sr- and Mn-doped $LaAlO_3$, which produced nine times more $H_2$ than $CeO_2$ when reduced at 1350 °C and re-oxidized at 1000 °C. [11] $La_{1-x}Sr_xMnO_3$ reduces to a greater extent than ceria, but its re-oxidation by steam is thermodynamically less favorable. This creates a trade-off between reduction extent and steam-to-$H_2$ conversion that is controlled by the Sr content. [12] Subsequent work tuned this balance through A-site Ca and small rare-earth cations [13] and B-site Al. [14] A second generation of materials combines a redox-active B-site cation with a stabilizing one, as in $BaCe_{0.25}Mn_{0.75}O_3$ (about three times $CeO_2$) [15], $SrTi_{0.5}Mn_{0.5}O_3$ (7.4 mL $g^{-1}$) [16] and $CaTi_{0.5}Mn_{0.5}O_3$ (10.0 mL $g^{-1}$ at 1350/1150 °C) [17], or introduces Ce as an additional redox-active cation [18-20]. More recently, compositionally complex and high-entropy perovskites have been explored. Examples include high-entropy manganites [21] for example $(La_{0.8}Sr_{0.2})(Mn_{0.2}Fe_{0.2}Co_{0.4}Al_{0.2})O_3$, which produced 395 µmol $g^{-1}$ and was stable over more than 50 cycles [22]. High-entropy oxides of other structure types have also been explored. In a direct comparison, high-entropy oxides outperformed ceria, but not La–Mn perovskites. [23]

### 1.1.2. Reduction with fuels

In CLHP, the oxygen carrier is reduced by a low-grade, and preferably biogenic, gaseous fuel that is converted fully to $CO_2$ and $H_2O$. After condensation of the water, the reduction step yields a $CO_2$ stream ready for capture, while the steam step yields $H_2$ free of carbon-containing products. Complete fuel conversion requires a counter-current fuel reactor. In a three-reactor configuration the carrier is finally re-oxidized by air, a highly exothermic reaction which restores its full oxygen capacity and supplies heat to the process. Iron oxides promoted with $La_{0.8}Sr_{0.2}FeO_3$ have reached 77 % steam-to-$H_2$ conversion at 930 °C, compared with at most 20 % for unpromoted ferrites [24], and Fe-doped $BaMnO_3$ reached 90 % [25]. Perovskite matrices also stabilize iron oxide against sintering [26].

Perovskite oxygen carriers have also been studied for chemical looping steam methane reforming (CLSMR), which uses the same steam-oxidation step and is therefore included in Table 1. CLSMR differs from CLHP in that the fuel is only partially oxidized, so the reduction step yields syngas rather than $CO_2$ ready for capture. $CH_4$ is the reductant, and the perovskite acts as a redox catalyst that converts it selectively to syngas while storing reduction capacity for the subsequent steam step, so that syngas and $H_2$ are co-produced. $LaFeO_3$-based carriers have been studied extensively for CLSMR. They yield syngas with $H_2/CO \approx 2$ in the reduction step and about 4–6.7 mmol $H_2$ $g^{-1}$ in the steam step for Co-, Sr- and Sr/Co-substituted compositions over 20–50 cycles. [27-29] Ce, Ni and Cu additions improve $CH_4$ activation and coke resistance, partly through reversible exsolution of metal nanoparticles, and give $H_2$ purities above 99.5 % [30]. Isothermal solar-driven operation has been demonstrated with $SrFeO_3$ at 950–1050 °C [31]. The $H_2$ yields in Table 1 are reported in different units (mL g−1, µmol g−1, mmol g−1, or as a multiple of $CeO_2$) and were obtained under different reduction and oxidation conditions and cycle durations and may therefore not be directly comparable.

*Table 1. Experimental studies of perovskite and perovskite-based oxygen carriers for hydrogen production by two-step thermochemical water splitting (thermal reduction), CLHP and CLSMR (fuel reduction). $T_{red}$ = reduction temperature; $T_{ox}$ = oxidation (water-splitting) temperature. $H_2$ yields are given as reported in the original studies. COG = coke-oven gas; n.r. = not reported.*

| Oxygen carrier | Reduction ($T_{red}$, atmosphere) | Oxidation ($T_{ox}$, oxidant) | $H_2$ yield / performance | Cycles | Ref. |
|---|---|---|---|---|---|
| a) ***Thermal reduction:*** *two-step thermochemical solar water splitting* | | | | | |
| Sr-, Mn-doped $LaAlO_3$ (SLMA) | 1350 °C | 1000 °C, $H_2O$ | 9 × $CeO_2$ ($H_2$)<br>6 × $CeO_2$ (CO) | ≥ 80 ($CO_2$ splitting) | [11] |
| $La_{1-x}Sr_xMnO_3$ (x = 0.3–0.4) | 1400 °C | 800 °C, $H_2O$ | ≈ 9 mL $g^{-1}$ (thermodynamic capacity, x = 0.4) Higher Sr content lowers steam conversion | n.r. | [12] |
| $La_{1-x}Ca_xMnO_3$; $Ln_{0.5}A_{0.5}MnO_3$ (best $Y_{0.5}Sr_{0.5}MnO_3$) | 1200 °C | 900 °C, $H_2O$ | Higher than $La_{0.5}Sr_{0.5}MnO_3$ and $CeO_2$ | n.r. | [13] |
| $La_{0.6}Sr_{0.4}Mn_{0.6}Al_{0.4}O_3$, $La_{0.6}Ca_{0.4}Mn_{0.6}Al_{0.4}O_3$ | 1200–1400 °C, low $pO_2$ | Isothermal 1240 °C or ΔT ≥ 200 °C, $CO_2$ | Reduction extent 5–13 × $CeO_2$; fuel yield up to 10 × $CeO_2$ (CO) | n.r. | [14] |
| $BaCe_{0.25}Mn_{0.75}O_3$ | 1250-1400 °C | 750-1000 °C 40% $H_2O$ | ≈ 3 × $CeO_2$ | n.r. | [15] |
| $SrTi_{0.5}Mn_{0.5}O_3$ | 1350 °C, $pO_2 \approx 10^{-5}$ atm | 1100 °C, $pH_2O$ = 0.4 atm | 7.4 mL $g^{-1}$; $H_2:O_2$ = 2:1 | no degradation | [16] |
| $CaTi_{0.5}Mn_{0.5}O_3$ | 1350 °C, Ar | 1150 °C, $H_2O$ | 10.0 ± 0.2 mL $g^{-1}$ per cycle | Stable to 1500 °C | [17] |
| Ce-substituted $(Ba,Sr)MnO_3$ | 1350-1400 °C, Ar | 850-1000 °C,40% $H_2O$ in Ar | 2–3 × $CeO_2$ | n.r. | [18] |
| $(Ce_xSr_{1-x})_{0.95}Ti_{0.5}Mn_{0.5}O_3$ (x ≤ 0.20) | 700–1400 °C (TGA) | $pO_2 = 1–10^{-4}$ bar | ΔH ≈ 238 kJ $mol^{-1}$<br>ΔS ≈ 141 J $mol^{-1}$ $K^{-1}$ (per mol O) | Phase-stable (x ≤ 0.15) | [20] |
| $Ca_{2/3}Ce_{1/3}Ti_{1/3}Mn_{2/3}O_3$ | - | - | Comparable to or higher than SLMA and $BaCe_{0.25}Mn_{0.75}O_3$ | Stable | [32] |
| $La_{1-x}Sr_xMnO_3$ reticulated porous foam | 1400 °C | n.r. | ≈ 200 µmol $g^{-1}$ per cycle | 50 (≈ 35 h) | [33] |
| $(La_{0.8}Sr_{0.2})(Mn_{0.2}Fe_{0.2}Co_{0.4}Al_{0.2})O_3$ | 1350 °C<br>$N_2$ | 800-1200 °C<br>40% $H_2O$ in $N_2$ | 395 ± 11 µmol $g^{-1}$ | > 50 | [22] |
| $(La_{1/6}Pr_{1/6}Nd_{1/6}Gd_{1/6}Sr_{1/6}Ba_{1/6})MnO_3$ | 1350 °C | 900, 1100 °C | ≈ 77.5 mmol $mol^{-1}$ oxide vs ≈ 53.5 for $La_{2/3}Sr_{1/3}MnO_3$ | 50 | [21] |
| ***b) Chemical looping hydrogen production*** *(CLHP): $H_2$ from the water-splitting step* | | | | | |
| Magnetite fines (Fe-oxide) | 800-975 °C, $CH_4$, CO, $H_2$ | 600-900 °C | 1.99 mmol $g^{-1}$ | 50 | [7, 34] |
| $Fe_2O_3$ in $La_{0.7}Sr_{0.3}FeO_3$ matrix | CO | 850 °C | 8 × $Fe_2O_3$ and 4 × $La_{0.7}Sr_{0.3}FeO_3$ alone | 25 | [26] |
| Ni-, Cu-doped $Ca_2Fe_2O_5$ (brownmillerite) | $CH_4$ | 850 °C | Higher steam-to-$H_2$ conversion than $Fe_2O_3$ | 15 | [35] |
| $La_{0.8}Sr_{0.2}FeO_3$-promoted $Fe_3O_4$ | $CH_4$ | 930 °C | 77 % $H_2O$-to-$H_2$ conversion (≤ 20 % for unpromoted ferrites) | 12 | [24] |
| $BaMn_{0.5}Fe_{0.5}O_3$ | $CH_4$ | 900 °C | 90 % steam-to-$H_2$ conversion | 18 | [25] |
| $LaNi_{0.07}Fe_{0.93}O_3$ | COG, 800 °C | 700-900 °C | +200% $LaFeO_3$; >99.7 % purity | 50 | [36] |
| $Ba_{0.5}Sr_{0.5}Co_{0.8}Mn_{0.2}O_3$ | $CH_4$, 850 °C | 850°C | Stable $H_2$ production for ≈ 200 min; $CH_4$ conversion up to 80 % | - | [37] |
| $La_{0.7}Sr_{0.3}FeO_3$ | 5 % CO/He, 850 °C | 850 °C, 1 % $H_2O$ | 0.82–0.90 mmol $g^{-1}$ (1.44 in 1st cycle) | Stable | [38] |
| $La_{0.6}Sr_{0.4}Ni_xFe_{1-x}O_3$ (x = 0, 0.1, 0.33) | CO, 900 °C | 900 °C, 8 % $H_2O$ | 0.21, 0.74 and 0.96 (x = 0, 0.1, 0.33) | n.r. | [39] |
| $Sr_2Fe_{1.5}Mo_{0.5}O_6$ | 5 % CO/Ar, 700 °C | 700 °C, ≈ 4 % $H_2O$ | ≈ 0.62 mmol $g^{-1}$ (≈ 0.19 for $SrFeO_3$) | n.r. | [40] |
| $(La_{1-x}Ca_x)MnO_3$ (x = 0–0.4) | $CH_4$ pulses, 900 °C | 900 °C, $H_2O$ pulses | 0.147–0.255 mmol $g^{-1}$, increasing with Ca content | n.r. | [41] |
| **c) *Chemical looping steam methane reforming*** (CLSMR): $CH_4$ partially oxidized to syngas, $H_2$ from the water-splitting step | | | | | |
| Three-dimensionally ordered macroporous (3DOM) $LaFeO_3$ | $CH_4$ | 850°C | Stable syngas and $H_2$ yields | 10 | [42] |
| $LaFe_{1-x}Co_xO_3$ x=0.1, 0.3, 0.5, 0.7, 1.0 | $CH_4$ | 850°C | ≈ 4 mmol $g^{-1}$ | 20 | [27] |
| $La_{2-x}Sr_xFeCoO_6$ double perovskite, x= 0, 0.2, 0.4, 0.6, 0.8, 1.0 | $CH_4$ | 850°C | ≈ 5.9 mmol $g^{-1}$ 93.8 % $H_2$ purity | 20 | [28] |
| $CeO_2$ / 3DOM $LaFeO_3$ | $CH_4$ | 800°C | 3.38 mmol $g^{-1}$ improved coke resistance | 30 | [43] |
| $La_{0.5}Sr_{0.5}FeO_3$ | COG, > 800 °C | 850 °C | 3.34 mmol $g^{-1}$ 99.85 % purity | 30 | [44] |
| $La_{0.95}Ce_{0.05}Ni_xFe_{1-x}O_3$ (x = 0.2, 0.5) | $CH_4$ | 900°C | > 99.5 % $H_2$ purity | 10 | [30] |
| $La_{0.8}Sr_{0.2}FeO_3$ | 10 % $CH_4$, 800 °C | 800 °C | 6.48–6.69 mmol $g^{-1}$ | 50 | [29] |
| $SrFeO_3$ foam (solar reactor, isothermal) | 5-30 % $CH_4$, 950–1050 °C | 1000 °C | 187-215 mL $g^{-1}$ | 12 | [31] |
| $LaMn_{0.5}Al_{0.5}O_3$ | $CH_4$ | Steam | 1.98 mmol $g^{-1}$ | n.r. | [45] |

### 1.1.3. Data-driven discovery

Despite this body of work, Table 1 shows that experimental studies have explored a limited set of compositions, mainly La-, Sr- and Ca-based manganites and ferrites substituted one or two cations at a time. Because the studies use different conditions, it remains unclear which compositions are best suited for CLHP, where the same oxygen carrier must be reduced by a fuel and still be re-oxidized by steam. The trade-off between reduction extent and steam-to-$H_2$ conversion [12] suggests that better carriers require combinations of several cations that tune the oxygen chemical potential, and such combinations are difficult to find by trial and error. The compositional space of substituted double and multi-component perovskites is too large to explore experimentally. Hence, high-throughput density functional theory (DFT) has been used to screen thousands of $ABO_3$ compositions for two-step thermochemical water splitting [8, 46, 47], which shares the steam-oxidation step with CLHP but reduces the carrier thermally rather than with a fuel, and to derive descriptors for oxygen-vacancy formation [48] and temperature-dependent thermodynamic properties of metal oxides [49], but its cost grows rapidly with compositional complexity. More recent studies therefore combine DFT with machine learning [10, 50] or with active learning [51, 52]. Experimental validation has been reported in only some of these studies, for example for $Ca_{2/3}Ce_{1/3}Ti_{1/3}Mn_{2/3}O_3$ [32] and $Ba_{0.875}Ca_{0.125}Zr_{0.875}Mn_{0.125}O_3$ [50] (Table 2). Active learning reduces computational effort and instead of enumerating the design space, it iteratively acquires the compositions expected to be most informative, based on the predicted performance and its uncertainty. [9] Using such a protocol, we previously showed that promising multicomponent perovskite oxygen carriers can be identified from a small fraction of the design space. [51, 52] These studies targeted the oxygen transfer capacity for chemical looping air separation and reforming, not hydrogen production.

*Table 2. Computational and data-driven screening studies of perovskite oxygen carriers for thermochemical water splitting and chemical looping.*

| Method | Search space | Outcome | Experimental validation | Ref. |
|---|---|---|---|---|
| High-throughput DFT | 5329 cubic and distorted $ABO_3$ perovskites | 139 candidates for water splitting, e.g. $CeCoO_3$ and $BiVO_3$ | No | [8] |
| DFT combined with experimental screening | Perovskite solid solutions | Redox thermodynamics for thermochemical applications | Yes | [46] |
| DFT | Ca–Ce–M–O perovskites (M = 3d metal) | $Ca_{0.5}Ce_{0.5}MnO_3$, $Ca_{0.5}Ce_{0.5}FeO_3$, $Ca_{0.5}Ce_{0.5}VO_3$ | No | [47] |
| Interpretable descriptor model (DFT-trained) | A = Ca, Sr, Ba, La, Ce; B = Ti, Ni | Oxygen-vacancy formation energy, MAE 0.45 eV; candidate $(Bi,Y)(Fe,Co)O_3$ | No | [48] |
| DFT + experiment | Ca–Ce–Ti–Mn perovskites | $Ca_{2/3}Ce_{1/3}Ti_{1/3}Mn_{2/3}O_3$ | Yes | [32] |
| High-throughput DFT + machine learning | 2401 $Sr_xA_{1-x}Fe_yB_{1-y}O_3$; ≈ 227,000 high-entropy perovskites | Equilibrium $pO_2$ tunable over ≈ 20 orders of magnitude | Yes | [10] |
| Machine learning + DFT + experiment + system simulation | Perovskite oxides | $Ba_{0.875}Ca_{0.125}Zr_{0.875}Mn_{0.125}O_3$; reduction up to 250 °C lower than $CeO_2$ | Yes | [50] |
| Active learning (Gaussian process) + CHGNet | High-entropy $ABO_3$ oxygen carriers | e.g. $Ca_{13}CoMgMn_9Ti_5Y_3O_{48}$; ≈ 0.056 g $O_2$ $g^{-1}$ oxygen transfer capacity, using ≈ 10 % of the training data | No | [51] |
| Autonomous active learning | High-entropy perovskite oxygen carriers | Candidates for CLAS and CLR | No | [52] |

## 1.2. Aim and Scope

In this work, we build on the autonomous active-learning (AL) protocol [51, 52] with the aim of designing perovskite oxygen carriers for CLHP. Compared with earlier frameworks, three extensions are introduced: i) hydrogen production capacity is evaluated under the coupled conditions of the fuel and steam reactors, ii) the compositional design space is enlarged from 14 to 22 metals, and iii) candidates are selected using multi-objective ranking that accounts for practical deployment constraints. The objective is to identify materials that combine favorable predicted hydrogen production capacity with potential for integration into a scalable chemical looping process. Hence, materials are evaluated not only based on predicted performance, but also with respect to practical considerations relevant to large-scale deployment, including raw material cost, supply resilience, and potential environmental and health impacts associated with handling the precursor oxides and products.

# 2. Method

## 2.1. Computational framework

The framework builds on the autonomous active-learning protocol of our previous work. [51, 52] Here it is extended as outlined in Section 1.2, with the hydrogen production capacity (HPC) under three-reactor conditions (FR, SR, AR) as the target property. Each AL cycle consists of five steps (Figure 2): (1) the training database, initialized with randomly generated compositions (Section 2.1.1), is used to (2) train a generative model, a Wasserstein autoencoder (WAE), of the composition space and a surrogate model (deep ensemble) of the vacancy-dependent Gibbs free energy, (3) new compositions are generated from the WAE latent space, (4) their HPC is predicted and compositions are acquired by Thompson sampling, and (5) the acquired compositions are evaluated with a machine-learning interatomic potential and phonon calculations and added to the database. The final candidates are selected after the AL runs by a separate multi-objective ranking of all evaluated compounds (Section 2.2).

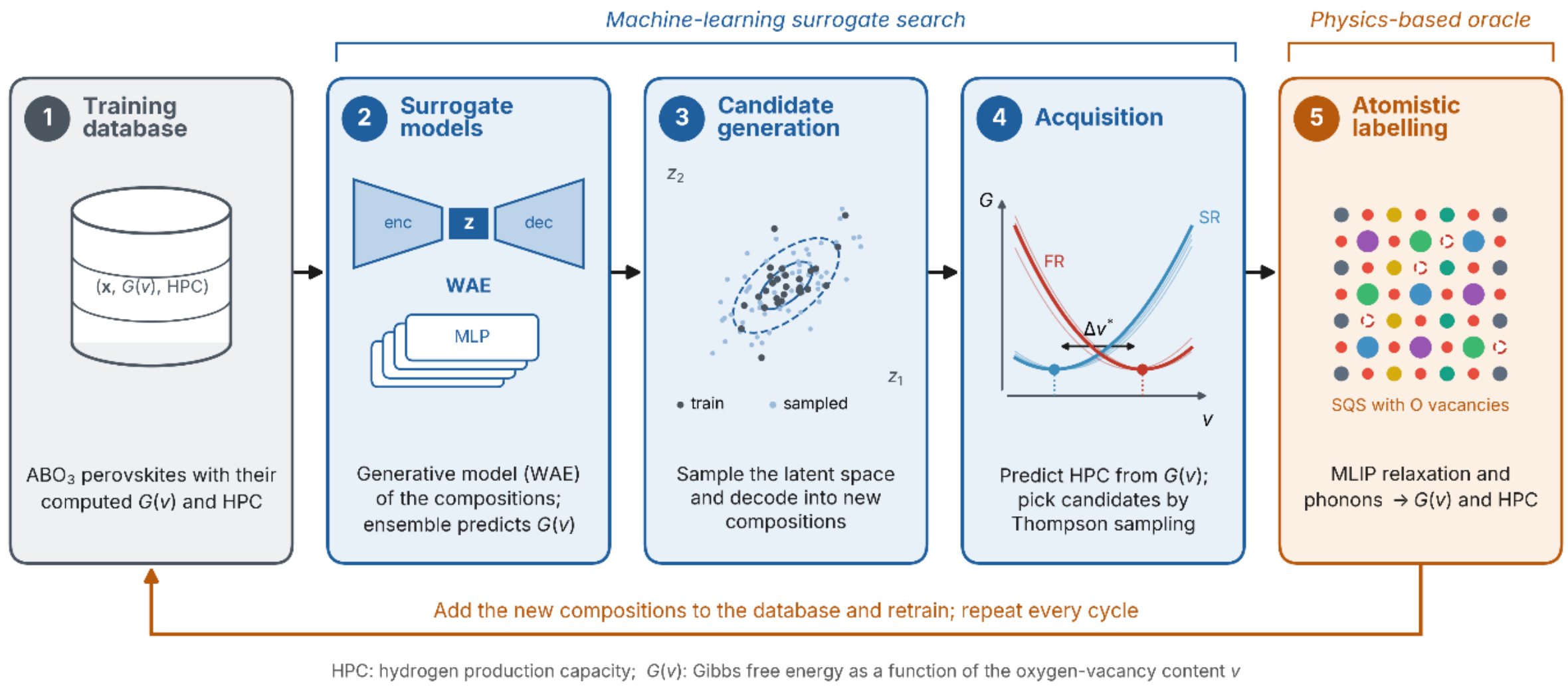

*Figure 2. Active-learning framework for discovering multi-component perovskite oxygen carriers for chemical looping hydrogen production. (1) The training database contains $ABO_3$ compositions modeled as 80-atom supercells, initialized with random compositions. (2) A Wasserstein autoencoder (WAE) maps compositions and their hydrogen production capacity (HPC) to a latent space, and a deep ensemble predicts the Gibbs free energy as a function of oxygen vacancy content, G(v), under fuel- (FR), steam- (SR) and air reactor (AR) conditions. (3) New compositions are generated by sampling the latent space and decoding. (4) HPC is obtained from the equilibrium vacancy contents $\delta_{FR}$ and $\delta_{SR}$, and compositions are acquired by Thompson sampling. (5) The acquired compositions are evaluated with CHGNet and harmonic phonon calculations on special quasi-random structures, added to the database, and the models are retrained.*

### 2.1.1. Database generation

Training data were generated using an 80-atom perovskite supercell (16 $ABO_3$ formula units). The A site was occupied by Ca, Ce, K, Mg, Y, Ba, La, Sm and Sr, and the B site by Al, Cr, Co, Cu, Fe, Mn, Ni, Sn, Ti, V, W, Zr and Zn. In total, 3903 compositions were generated by randomly choosing the elements and their fractions on each site, each containing two to seven distinct metallic elements, with one to six elements on each sublattice. Compositions were filtered for i) charge neutrality and ii) geometric stability, using the Bartel tolerance factor $\tau \leq 4.18$. [53] After filtering, 2103 compositions remained and formed the starting set (cycle 0), shared by all AL runs.

For a given composition, the arrangement of the cations over the A and B sites, i.e. the chemical disorder, was represented by special quasi-random structures (SQS) [54], which mimic a random distribution of cations within the finite supercell, constructed with the ICET package [55]. To capture temperature- and oxygen-partial-pressure-dependent structural effects, SQS were generated for both the cubic and the orthorhombic perovskite polymorph, each with 0 to 8 oxygen vacancies per 80-atom supercell ($\delta$ = 0 to 0.5 in steps of 1/16), i.e. nine structures per polymorph. One brownmillerite structure ($A_2B_2O_5$, $\delta$ = 0.5) with ordered vacancies was added, giving 19 structures per composition. In total, for each composition, thermodynamic properties were evaluated for 19 structures, and vacancy formation energies were determined for the most stable configuration.

### 2.1.2. Property evaluation

The redox properties of the perovskites were quantified in terms of oxygen vacancy formation free energies under reactor-specific conditions [49]. These energies determine the ability of a material to reversibly exchange lattice oxygen and thereby produce hydrogen (Section S1, Supplementary Information, SI). Energies and forces were computed with the machine learning interatomic potential CHGNet [56], which gives near-DFT accuracy at reduced computational cost. Structures were relaxed with the Broyden–Fletcher–Goldfarb–Shanno (BFGS) algorithm implemented in the Atomic Simulation Environment (ASE) [57], with forces converged to below 0.001 eV $Å^{-1}$. Vibrational contributions to the free energy were obtained using PHONOPY [58] from finite-displacement calculations (displacement amplitude 0.02 Å) in supercells at least 12.5 Å in each crystallographic direction. The Brillouin zone was sampled with approximately $10^6$ q-points per atom.

#### 2.1.2.1. Hydrogen production capacity (HPC)

The amount of hydrogen an OC can produce depends on how much its oxygen non-stoichiometry changes between reactors. The temperature and gas composition of each reactor fix the oxygen chemical potential, $\mu_O$, and hence the equilibrium vacancy concentration. $\mu_O$ was calculated for the FR, SR and AR from their respective temperatures and gas compositions (Table S1, SI), using temperature-dependent thermochemical data for the gas-phase reference species from the NIST-JANAF tables [59]. Combining these $\mu_O$ values with the calculated free energies gives the Gibbs free energy of each composition as a function of vacancy content G(v) under each reactor condition. The equilibrium non-stoichiometry was obtained by fitting a quadratic function to G at the eight vacancy contents predicted by the surrogate (0–8 vacancies per supercell) and taking the position of its minimum. This gives a continuous estimate of vacancies (directly correlated to $\delta$) rather than the discrete, most stable sampled vacancy count. The sensitivity of the HPC to the assumed SR gas ratio is shown in Figure S1 (SI).

The HPC follows from the change in non-stoichiometry between the FR and the SR, $\Delta\delta = \delta_{FR} - \delta_{SR}$ (see Section S1 of the SI for details). This difference is the amount of oxygen the carrier takes up from $H_2O$ in the SR, which equals the amount of $H_2$ produced per formula unit. The HPC was normalized to the molar mass of the oxidized perovskite:

$$HPC = 1000 \frac{\Delta\delta}{M_{ABO_3}} \; [mmol\ H_2\ g^{-1}] \qquad 1$$

### 2.1.3. Database update

Following property evaluation, new compositions were appended to the database, storing composition vectors, Gibbs free energies and the associated hydrogen production capacity. The generative and surrogate models were retrained each cycle to incorporate newly acquired information.

### 2.1.4. Deep ensemble

The surrogate model was employed to approximate the composition–property landscape and quantify predictive uncertainty. The deep ensemble consisted of 16 independently trained fully connected neural networks, each comprising two hidden layers with 64 units, SiLU activation functions, and dropout (rate 0.25). The input is the composition vector. Each network outputs a mean and a variance for the Gibbs free energy at each vacancy content in the FR and the SR, so that aleatoric and epistemic uncertainty can be estimated separately. [60] Targets were z-score normalized. Each member was trained by minimizing the Gaussian negative log-likelihood with the Adam optimizer (learning rate $1 \times 10^{-4}$, batch size 64, at most 500 epochs). Randomized maximum-a-posteriori sampling was applied using anchor regularization and target perturbation [61] to approximate posterior sampling and enable Thompson sampling [62]. Noise perturbations scaled with vacancy concentration to reflect increased uncertainty at higher non-stoichiometries. Details on architecture and hyperparameters are presented in Table S2 of the SI.

### 2.1.5. Generative model: Wasserstein Autoencoder

New compositions were generated with a Wasserstein autoencoder (WAE) [63] with an eight-dimensional latent space. Besides the composition decoder, the WAE has a regression head that predicts the HPC from the latent vector, so that the latent space is organized by both composition and performance. A maximum-mean-discrepancy penalty (weight 0.01, multi-scale inverse multiquadric kernel between the encoded latent vectors and samples from N(0, I)) aligns the distribution of latent codes with a standard normal prior. The encoder and decoder use dropout (rate 0.1) and weight decay ($1 \times 10^{-5}$). The WAE was first trained for 1500 epochs (Adam, learning rate $1 \times 10^{-3}$, batch size 32). In each subsequent cycle it was trained further, starting from the previous weights, for 500 epochs at a learning rate of $2 \times 10^{-4}$. Further details are given in Table S3 of the SI.

### 2.1.6. Sampling and acquisition

In each cycle, a single Gaussian was fitted to the latent codes of the current database, and 25 000 latent vectors were drawn from it and decoded into compositions. The decoded site fractions were rounded to integer cation counts on the 16 A and 16 B sites of the supercell. The ensemble then predicted G(v) for every composition. The HPC of each composition was derived from these predictions as in Section 2.1.2.1, separately for each ensemble member.

New compositions were acquired by Thompson sampling [62]. For each composition, one of the 16 members was drawn at random, and the HPC predicted by that member was used as the acquisition value. The ten compositions with the highest values were acquired for full property evaluation and added to the database. Because each member corresponds to an approximate posterior sample, a composition is acquired when at least one plausible realization of the surrogate predicts a high HPC. This balances exploitation of high-performing regions against exploration of uncertain ones.

To assess the sensitivity of the search to its stochastic elements, seven independent runs of 50 cycles were performed. All runs used identical settings and started from the same starting set (cycle 0) but with different random seeds. The runs are labeled r1–r7. The seed controls all random elements of the AL framework including the initialization of both models, the choice of the 15 % of compounds held out for validation in each cycle, the latent-space samples and the Thompson draws. The starting dataset in cycle 0 (Section 2.1.1) is the same in all runs. Because each cycle's acquisitions enter the next cycle's training data, these differences accumulate, and runs can follow different paths through the composition space.

### 2.1.7. Data visualization

To follow how the acquired compositions are distributed in the learned representation, we projected the latent space to two dimensions with Uniform Manifold Approximation and Projection (UMAP) [64]. Each composition was encoded with the WAE encoder of run r6, the run that reached the highest HPC, as trained in AL cycle 50. The encoder input is the same as in training, the HPC, standardized with the mean and standard deviation of the starting set and r6, and the cation site fractions of the 22 elements, each scaled by its sublattice factor. This gives a 23-dimensional input, and an 8-dimensional latent vector. All compositions with an HPC label from cycles 0-50 were included. The UMAP model was fitted to the latent vectors of the starting set and of the compositions acquired by r6, using 15 nearest neighbors, a minimum distance of 0.1, the Euclidean metric and a random seed of 42. The model was then held fixed, and the compositions of the other six runs were placed on it with the fitted transform. All runs therefore share one coordinate system. The UMAP does not preserve distances, and the projection is used qualitatively, to show where each run's compositions lie relative to the starting set and each other.

## 2.2. Candidate selection

After the AL runs were completed, candidate materials were selected from all evaluated compounds using a multi-objective ranking that balances the predicted performance against practical considerations for large-scale deployment. Four criteria were considered: i) HPC, ii) raw material price ($ $kg^{-1}$, 2018 values) [65], iii) a supply resilience (Res) index, and iv) an environmental and health (EH) index.

The Res score of each element is $1 - R/R_{max}$, where R is the supply risk indicator from the 2023 EU criticality assessment [66], which underlies the Critical Raw Materials Act (Regulation (EU) 2024/1252) [67]. The EH index of each element is taken from the hazard classifications of its binary oxides under the EU CLP Regulation (EC) No 1272/2008, using its most hazardous oxide. It ranges from 1 (no hazard classification) to 0 (most severe profile). Full details are given in Section S3 and Tables S4–S6 of the SI.

The candidate pool consisted of the starting set and the compounds acquired during the AL runs, restricted to charge-neutral compositions with $\tau \leq 4.18$ and HPC > 0. Because the criteria have different units and ranges, each was converted to a percentile rank so that 100 is always best. The combined score is the weighted sum of these percentiles, with weights 0.6 / 0.1 / 0.15 / 0.15 for HPC / Price / Res and EH, respectively.

The highest combined scores are concentrated in closely related compositions, so a selection by score alone would return similar candidates. 20 candidates were therefore selected by maximal marginal relevance (MMR) [68], considering only compounds with HPC ≥ 1.0 mmol $H_2$ $g^{-1}$. Starting from an empty set S, the compound i that maximizes

$$\lambda s_i - (1 - \lambda) \cdot max_{j \in S}\ sim(i, j) \qquad 2$$

is added to S, and this is repeated until 20 compounds have been selected. Here $s_i$ is the combined score of compound i, min–max normalized to [0, 1]. sim(i, j) is the Jaccard index of the B-site element sets of i and j, counting only elements that occupy at least two of the 16 B sites so that single-site dopants are ignored. For example, sim = 2/3 for Sn–Ti vs Sn–Ti–V and sim = 1/3 for Sn–Ti vs Fe–Sn. Compounds with the same B-site element set as an earlier pick were excluded, so each B-site chemistry is represented at most once. Here, λ = 0.5, which weights score and diversity equally.

# 3. Results and discussion

## 3.1. Exploring high-performing multicomponent oxides

Active learning raised the hydrogen production capacity (HPC) of the acquired compounds well above that of the starting set within the first few cycles and did so in all seven runs. Figure 3 shows how the mean HPC evolves over 50 cycles in seven independent runs. The mean HPC of the AL-acquired compounds rose from 0.41 to 1.56 mmol $H_2$ $g^{-1}$ in the first cycle. It then increased further and reached 2.05 mmol $H_2$ $g^{-1}$ in cycle 50 (Figure 3). All runs followed the same trend, although their levels differed: in cycles 46–50 the run means ranged from 1.82 to 2.42 mmol $H_2$ $g^{-1}$. In every cycle, the AL-acquired compounds had higher HPC than the starting set.

The highest HPC found in each run ranged from 2.65 to 2.99 mmol $H_2$ $g^{-1}$. r1 reached its maximum (2.70 mmol $H_2$ $g^{-1}$) as early as cycle 5, but the other runs kept improving until late in the search: r2 from 2.50 to 2.82 (cycle 44), r3 from 2.34 to 2.66 (cycle 41), r4 from 2.35 to 2.65 (cycle 46), r5 from 2.32 to 2.85 (cycle 50), r6 from 2.46 to 2.99 (cycle 34) and r7 from 2.50 to 2.78 mmol $H_2$ $g^{-1}$ (cycle 50). Every run also acquired compounds with HPC = 0, from 11 % in r3 to 34 % in r5, which shows that the search kept exploring regions outside the high-performing domain.

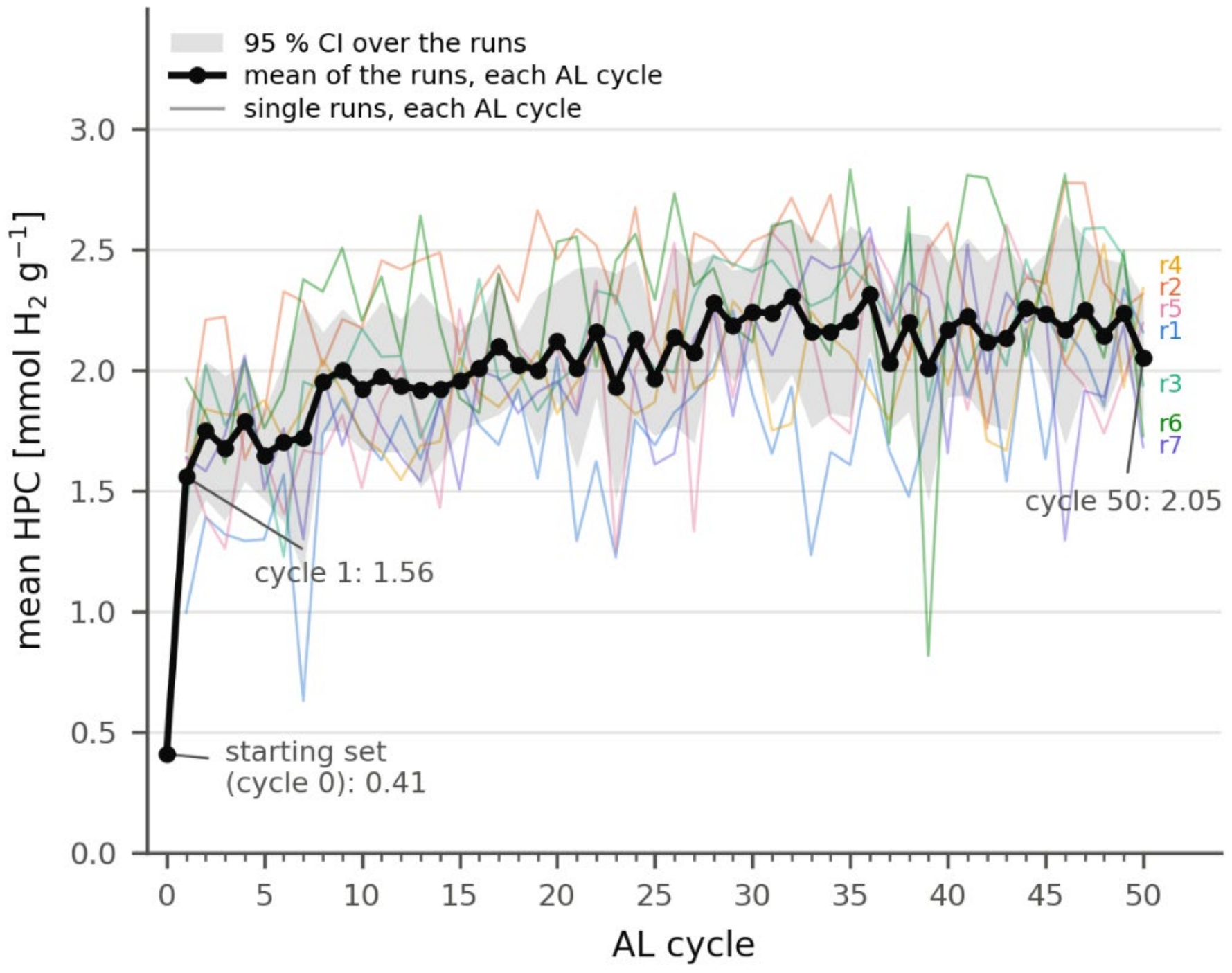

*Figure 3. Mean hydrogen production capacity (HPC) of the compounds acquired by active learning in seven independent runs (r1–r7), which differ only in their random seed. The black line is the mean over the seven runs for each single cycle. The shaded band is the 95 % confidence interval over the runs. The point at cycle 0 is the starting set (2103 compounds, shared by all runs; mean 0.41 mmol $H_2$ $g^{-1}$). The mean rises to 1.56 mmol $H_2$ $g^{-1}$ in the first cycle and reaches 2.05 mmol $H_2$ $g^{-1}$ in cycle 50.*

To compare the runs, all AL-acquired compositions were encoded with a single WAE and projected with UMAP into a common two-dimensional map of the latent space, in which nearby points have similar compositions and HPC but distances between distant clusters are not quantitative (Figure 4, details provided in Section S2.4 and Figure S2 of the SI). Although the runs share no acquired compositions, they converge on the same part of the compositional space. The compounds of every run cluster in the top right region of the map, with a mean HPC of 2.1 mmol $H_2$ $g^{-1}$ against 0.41 mmol $H_2$ $g^{-1}$ for the starting set. Chemically, this region is dominated by Sn on the B site (on average 9 of 16 B-site cations), combined with Ti and with the A-site cations K, Mg, Ce and Y (Figure S2). It contains the best compounds of every run, up to 2.99 mmol $H_2$ $g^{-1}$. Only 93 of the 2103 starting compounds (4 %) fall in the same region, with a mean HPC of 0.52 mmol $H_2$ $g^{-1}$, so the runs did not just pick out the best-known compositions but moved into a sparsely sampled part of the space and filled it with compounds four times better on average. Note, however, that Sn alone does not imply a high HPC; these compounds cluster on the map partly because HPC is part of the WAE input. Only one run deviated from this pattern. In addition to the Sn-rich compositions, r1 explored Mn-rich perovskites (Mn on ≈ 8 of the 16 B sites, with Ce, Mg and Y on the A site). Of the 166 Mn-rich compounds acquired by AL, 160 come from r1. Mn-rich perovskites therefore form a second solution that the search finds only occasionally from the same starting point, but on average these compositions exhibit a lower HPC (mean 1.77, best 2.60 mmol $H_2$ $g^{-1}$) compared to the Sn-based ones (up to 2.99 mmol $H_2$ $g^{-1}$).

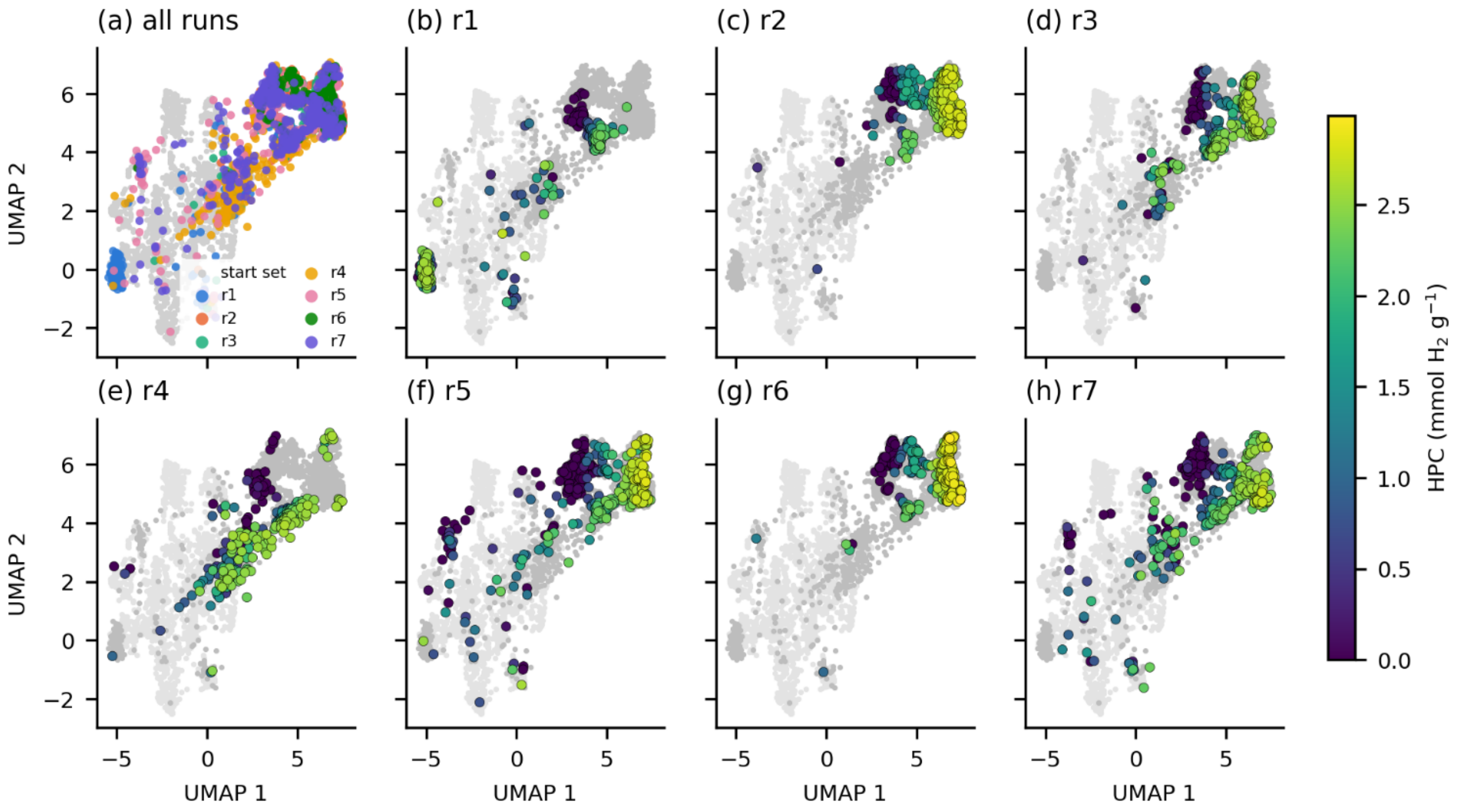

*Figure 4. Compositions explored by the seven active-learning runs, shown in a common two-dimensional map of the latent space. All compounds were encoded with the Wasserstein autoencoder (WAE) of run r6, the run that reached the highest HPC, as trained in AL cycle 50. A UMAP projection (15 neighbors, minimum distance 0.1, random seed 42) was fitted to the starting set and the r6 compounds, then held fixed while the compounds of remaining runs were projected onto it. All runs therefore share the same coordinates. (a) Starting set (light grey, 2103 compounds) and the compounds added by all runs, (b–h) Compounds added by each run, colored by HPC, against the starting set (light grey) and the compounds of the other runs (dark grey).*

## 3.2. Chemical domains for oxygen carrier design

Figure 5 compares the site fraction of each element in the compounds acquired by AL with that in the starting set, where a ratio above 1 indicates enrichment. On the B-site, Sn was strongly enriched in all seven runs (6.6–8.5, 7.5 pooled over runs). Ti was enriched in five runs (r6: 5.0, r2: 4.6, r5: 3.5, r7: 3.3, r3: 2.7) but depleted in r1 and r4 (0.2–0.3), and Al was enriched in r3, r4 and r5 (1.7–4.1). Mn was enriched only in r1 (3.4), where it was accompanied by a weak enrichment of Zn (1.6). Cu was enriched only in r7 (1.8). V, Fe, Co, Ni, Cr, W and Zr were depleted in all runs, apart from V in r2, which stayed close to its initial fraction (0.9). On the A-site, K was enriched in every run (2.2–5.9, 3.5 pooled), followed by Mg, which was enriched in six runs (2.8–4.1) but depleted in r3 (0.3). Ce was enriched in r1, r3, r4 and r7 (1.3–2.2) and depleted in r2, r5 and r6 (0.4–0.6). Y was enriched in six runs (1.3–1.8) but almost absent in r4 (0.07). Ca remained close to or below its initial fraction (0.5–1.5), and Ba, La and Sm were depleted in all runs, as was Sr except in r4 (1.1).

The elemental enrichment points to one dominant chemical domain, namely Sn-rich B-sites combined with K/Mg and Ce/Y on the A-site. The highest-performing compositions, however, differed between runs. In r2, r3, r5, r6 and r7, all ten compositions with the highest HPC (2.61–2.99 mmol $H_2$ $g^{-1}$) contained both Sn and Ti, in similar amounts (on average 7/16 B-sites). In r4, Sn was combined with Al and V instead of Ti, and all of the top ten contained Ce and Mg. r1 converged on a different domain where eight of its top ten compositions (2.55–2.70 mmol $H_2$ $g^{-1}$) were Mn-rich (B-site Mn fraction ≥ 0.5), only one contained Sn, and all contained Ce and Mg, mostly together with Y.

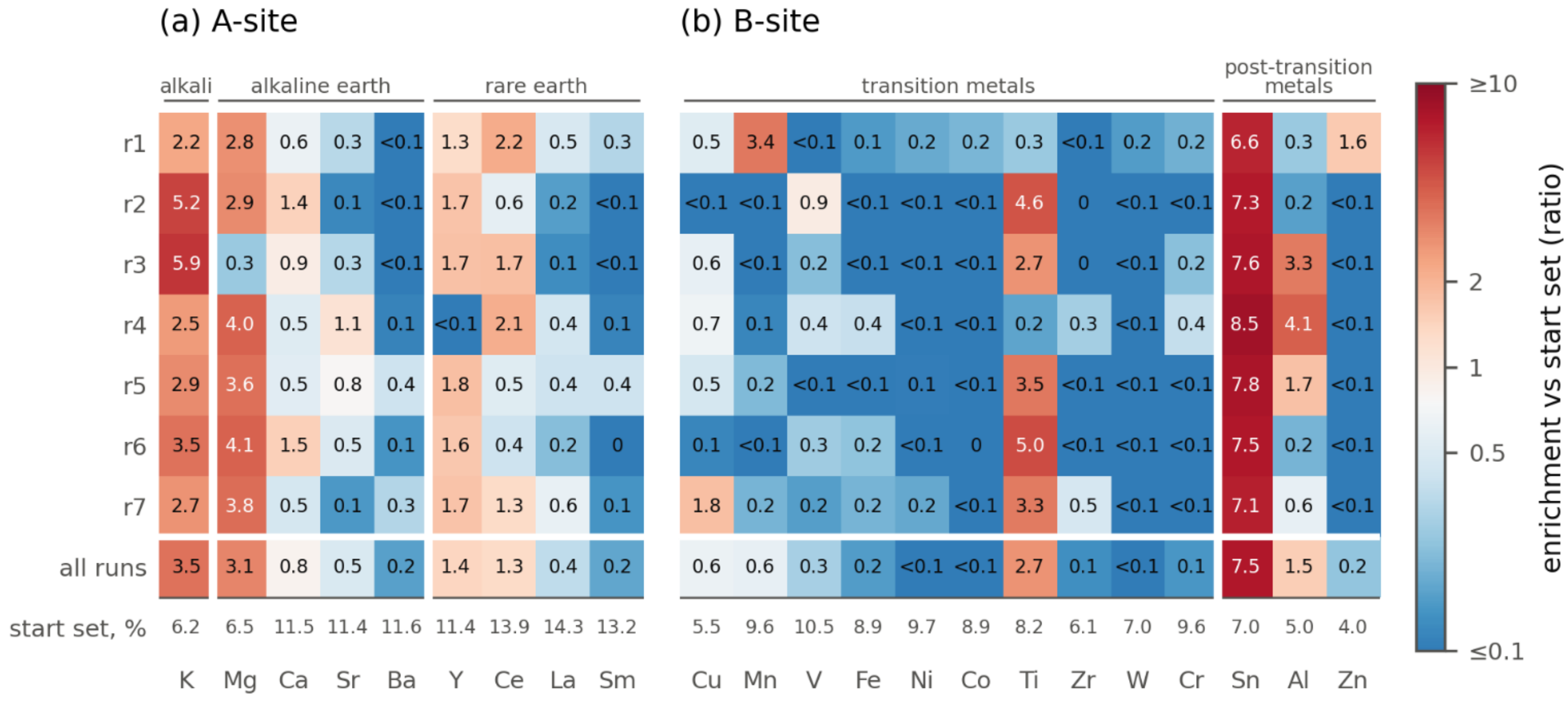

*Figure 5. Element enrichment on the (a) A-site and (b) B-site in the compounds acquired by active learning, relative to the starting set. Each cell gives the mean site fraction of an element in the AL-acquired compounds of one run (r1–r7) or of all runs combined, divided by the mean site fraction in the starting set; 1 = no change, "<0.1" means a ratio below 0.1. The color scale is logarithmic and clipped at 0.1 and 10. The bottom row gives the mean site fraction (%) of each element in the starting set. Columns are grouped by chemical family.*

## 3.3. Linking composition to redox capacity

A high HPC requires a large $\Delta\delta$, i.e., a composition that is reduced under FR conditions and re-oxidized by steam. In 99 % of the 673 compositions with HPC ≥ 2.5 mmol $H_2$ $g^{-1}$, the equilibrium swings between fully oxidized in the SR ($\delta = 0$) and the largest vacancy content in the model ($\delta = 0.5$) in the FR, hereafter a full swing. Whether a composition reaches the full swing depends on the combination of A- and B-site cations. Of the 3006 compositions with at least 4 Sn per cell (starting set and AL), 47 % reach it, 28 % reach a partial swing, whereas 17 % remain oxidized and 8 % remain reduced in both reactors and therefore have no capacity. Compositions with a full swing contain twice as much K as those that remain oxidized (4.2 vs 2.1 per cell) and less La and Sr (0.7 vs 2.0 and 0.7 vs 1.5). As an example, the reduction midpoint ($\delta = 0.25$) of Sn-rich compositions shifts towards more oxidizing conditions with increasing K, but is almost unchanged by Ti (Figure S3, SI). The direction of this shift matters because the redox window is not centered on the typical composition. At 800 °C the median midpoint lies 57 kJ $mol^{-1}$ $O_2$ below the SR gas line, but at 900 °C only 23 kJ $mol^{-1}$ $O_2$ above the FR gas line. Accordingly, compositions that miss the redox window almost always do so because they are too hard to reduce (21 %), rarely because they are too easy to reduce (2 %). An element that raises the midpoint therefore increases the chance of a full swing. To separate the contribution of each element from the cations it is combined with, we fitted the midpoint of all 2712 AL compositions against all A- and B-site element counts (Figure 6).

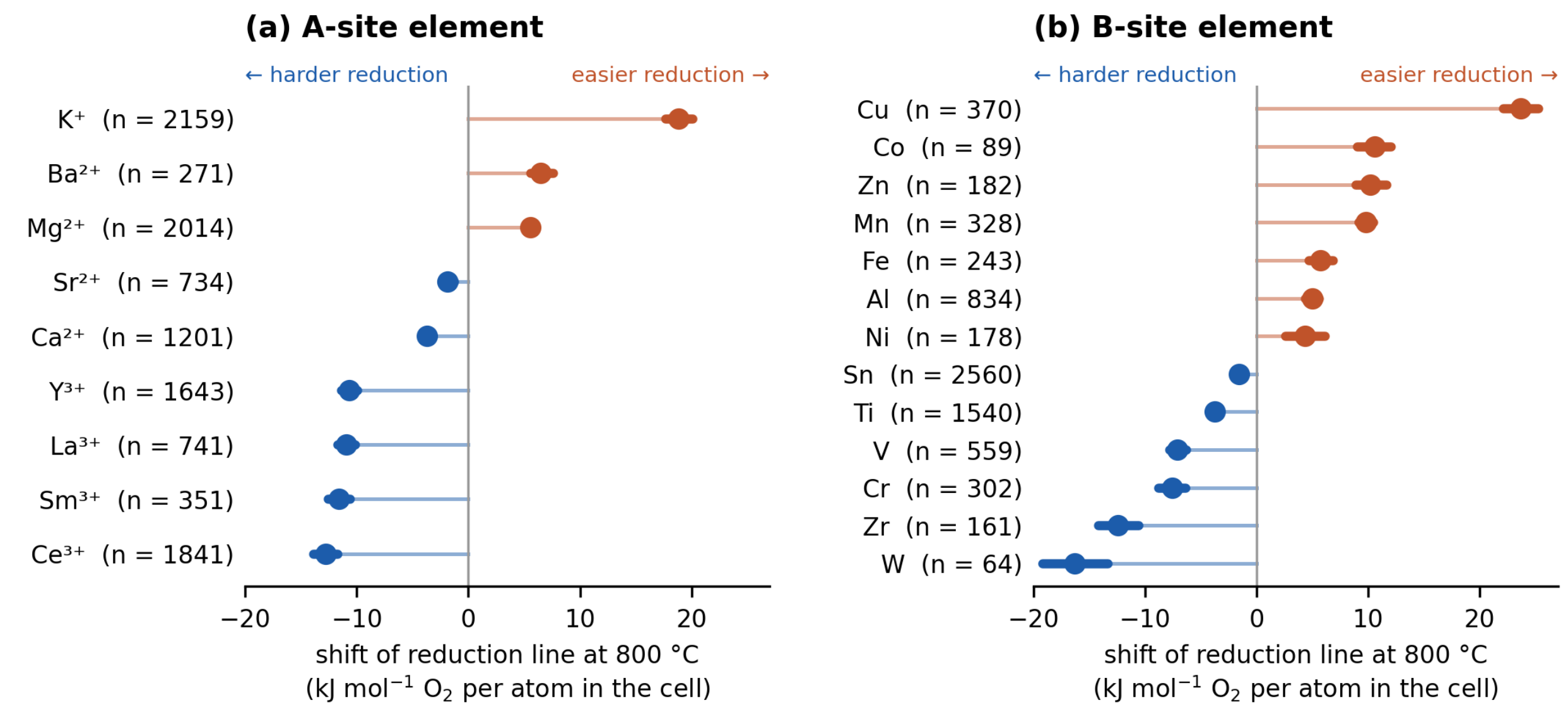

*Figure 6. Effect of (a) A-site and (b) B-site elements on the reduction midpoint (δ = 0.25) at 800 °C, from a joint fit of the midpoint of 2712 AL compositions against all A- and B-site element counts. Each dot gives the shift, in kJ mol $^{-1}$ $O_2$ per atom in the 80-atom cell (16 A-sites, 16 B-sites, 48 O), when one average cation of that site is replaced by the element with all other elements unchanged; bars give the 95 % confidence interval. The zero line is the average cation of each site over the 2712 compositions, values are relative to this average. Positive values (orange) shift the midpoint towards more oxidizing conditions than the average cation, i.e. the oxide is reduced more easily; negative values (blue) make it harder to reduce. n is the number of compositions that contain the element.*

The zero line in Figure 6 is the average cation of each site, and hence the sign of values close to zero depends on this reference, whereas the differences between elements do not. On the A-site, the shift follows the cation charge (Figure 6a). $K^+$ is associated with the largest upward shift of the midpoint (+19 kJ mol$^{-1}$ $O_2$ per atom), although this estimate may partly reflect its frequent co-occurrence with Sn, Ti and Mg in the AL compositions. The divalent cations change it by −4 to +7 kJ mol$^{-1}$ $O_2$, whereas the trivalent cations (Y, La, Sm, Ce) lower it by 11–13 kJ mol$^{-1}$ $O_2$. This agrees with the defect chemistry of acceptor-doped perovskites such as Sr-doped $LaFeO_3$ and $LaCoO_3$, in which a lower A-site charge favors oxygen vacancy formation [69, 70]. On the B-site (Figure 6b), Sn (−2 kJ mol$^{-1}$ $O_2$) and Ti (−4 kJ mol$^{-1}$ $O_2$) lie close to the B-site average and thus form a stable background whose redox behavior the A-site cations tune. Transition metals such as Cu (+24 kJ mol$^{-1}$ $O_2$), Co and Mn (both about +10 kJ mol$^{-1}$ $O_2$), Fe and Ni (+4-6 kJ mol$^{-1}$ $O_2$) raise the midpoint, as does Zn (+10 kJ mol$^{-1}$ $O_2$) while W, Zr, Cr and V lower it.

In full swing, eight oxygen atoms are exchanged per 80-atom cell, and the HPC equals 8000/M mmol $H_2$ g$^{-1}$, with M the molar mass of the cell in g mol$^{-1}$. Once a full swing is reached, the HPC can only be increased by lowering M. This explains the incorporation of Ti, a light, redox-inactive tetravalent B-site cation that dilutes the heavier Sn without changing the redox behavior. Among Sn-rich compositions with a full swing, the mean HPC increases from 2.21 mmol $H_2$ g$^{-1}$ without Ti to 2.71 mmol $H_2$ g$^{-1}$ with 9–16 Ti per cell, as the cell mass decreases from 3640 to 2955 g mol$^{-1}$. K and Mg contribute in the same way on the A-site, and their low valence is balanced by the tetravalent B-site cations in the oxidized perovskite.

The redox window depends on the gas conditions assumed in the two reactors. To test this, the HPC of the 2103 compounds of the starting set was recalculated for $H_2O/H_2$ ratios between 0.1 and $10^6$ in the SR, with the FR kept at the AL conditions (Figure S1, SI). A more oxidizing SR also re-oxidizes compositions that are more easily reduced, which at lower ratios remain reduced in both reactors. Accordingly, the top 10 % of compounds shift towards Mn-rich B-sites with increasing $H_2O/H_2$, and the Mn enrichment rises from about 1.1 at $H_2O/H_2$ = 4 to about 2.5 at $10^6$. Figure S1c shows how robust the top-10 % ranking is to the assumed SR conditions, expressed as the share of the top 10 % (210 compounds) at each ratio that is also in the top 10 % at $H_2O/H_2$ = 4. Such oxidizing conditions are, however, of little practical relevance, since $H_2O/H_2 = 10^6$ corresponds to an equilibrium steam conversion of only about 0.0001 %. The Sn-rich domain is therefore favored under the practically relevant conditions used here ($H_2O/H_2$ = 4, $CO_2/CO$ = 1/9), which balance capacity against steam conversion.

Several limitations of the approach should be noted. First, the vacancy content is restricted to $0 \le \delta \le 0.5$ (0–8 vacancies per 80-atom cell). Compositions that would reduce further are capped at δ = 0.5, and 99 % of the compositions with HPC ≥ 2.5 mmol $H_2$ $g^{-1}$ swing across this entire range. For these compositions the predicted HPC is an upper bound set by the model, 8000/M mmol $H_2$ $g^{-1}$, and their ranking is governed mainly by the molar mass of the cell rather than by differences in redox thermodynamics; the enrichment of light cations such as Ti, K and Mg partly reflects this. Thus, materials that exchange more oxygen through deeper reduction or phase transitions, such as $SrFeO_3$ or iron oxides, are not credited with this additional capacity. Second, the HPC is an equilibrium quantity, and reaction kinetics, gas conversion, attrition, carbon deposition and degradation over repeated cycles, which determine performance in a real reactor, are not considered. The candidates should hence be regarded as ranked suggestions for experimental testing with promising redox properties for CLHP.

### 3.4. Multi-objective candidate ranking

After the seven AL runs were completed, candidates were selected by a multi-objective ranking of the 3108 compounds with HPC > 0 (Section 2.2). Ranked by the combined score alone, the 20 best compounds all combine Sn and Ti on the B-site (Sn-Ti, Fe-Sn-Ti), so a selection by score would give similar candidates. Figure 7 therefore shows 20 chemically diverse candidates selected by MMR (Section 2.2), and their rank for each criterion and for the combined score. The six highest-ranked candidates all contain both Sn and Ti. The Sn–Ti candidate $(Ce_3K_4Mg_8Sr)(Sn_6Ti_{10})O_{48}$ has both the best combined rank and the highest HPC among the candidates (2.84 mmol $H_2$ $g^{-1}$, HPC rank 28). 15 of the 20 candidates contain Sn on the B-site, the other five combine Cr–Fe, Al–Mn, Fe–Mn, Cr–Mn and Mn–Zn.

The individual rankings reveal clear trade-offs between performance and practical considerations. The Cr–Fe candidate $(Ce_7Mg_8Sm)(Cr_6Fe_{10})O_{48}$ combines high HPC (rank 233) and low price (rank 28) with a poor EH rank (2491). All five Mn-rich candidates rank poorly for supply resilience: the Mn–Zn candidate $(Y_{16})(CuMn_{13}Zn_2)O_{48}$ lies at the 3rd percentile, followed by the Al–Mn (7th), Mn–Sn-Ti (10th), Cr–Mn (15th) and Fe–Mn (20th) candidates, even though the Fe–Mn candidate $(Ce_{10}Mg_6)(Fe_8Mn_8)O_{48}$ has the third-best price rank of all compounds. Similarly, the Cu–Sn–V candidate ranks low for EH impact (2317), and the Sn–W–Zn and Mn–Zn candidates rank low for price (2073 and 2449). The Al–Sn–Ti candidate $(Ca_2K_4Mg_4Y_6)(Al_2Sn_7Ti_7)O_{48}$ has the best EH rank among the candidates (142) at a high HPC rank (62) but a low-price rank (1819), whereas the Fe–Sn–Ti candidate $(Ca_4KMg_6Sr_2Y_3)(Fe_3Sn_7Ti_6)O_{48}$ ranks within the best 250 compounds for HPC, Res and EH and reaches the second-best combined rank among the candidates (9). The combined ranking therefore provides a compromise between performance and the practical requirements of large-scale deployment, and the combined ranks of the 20 candidates range from 1 to 1226.

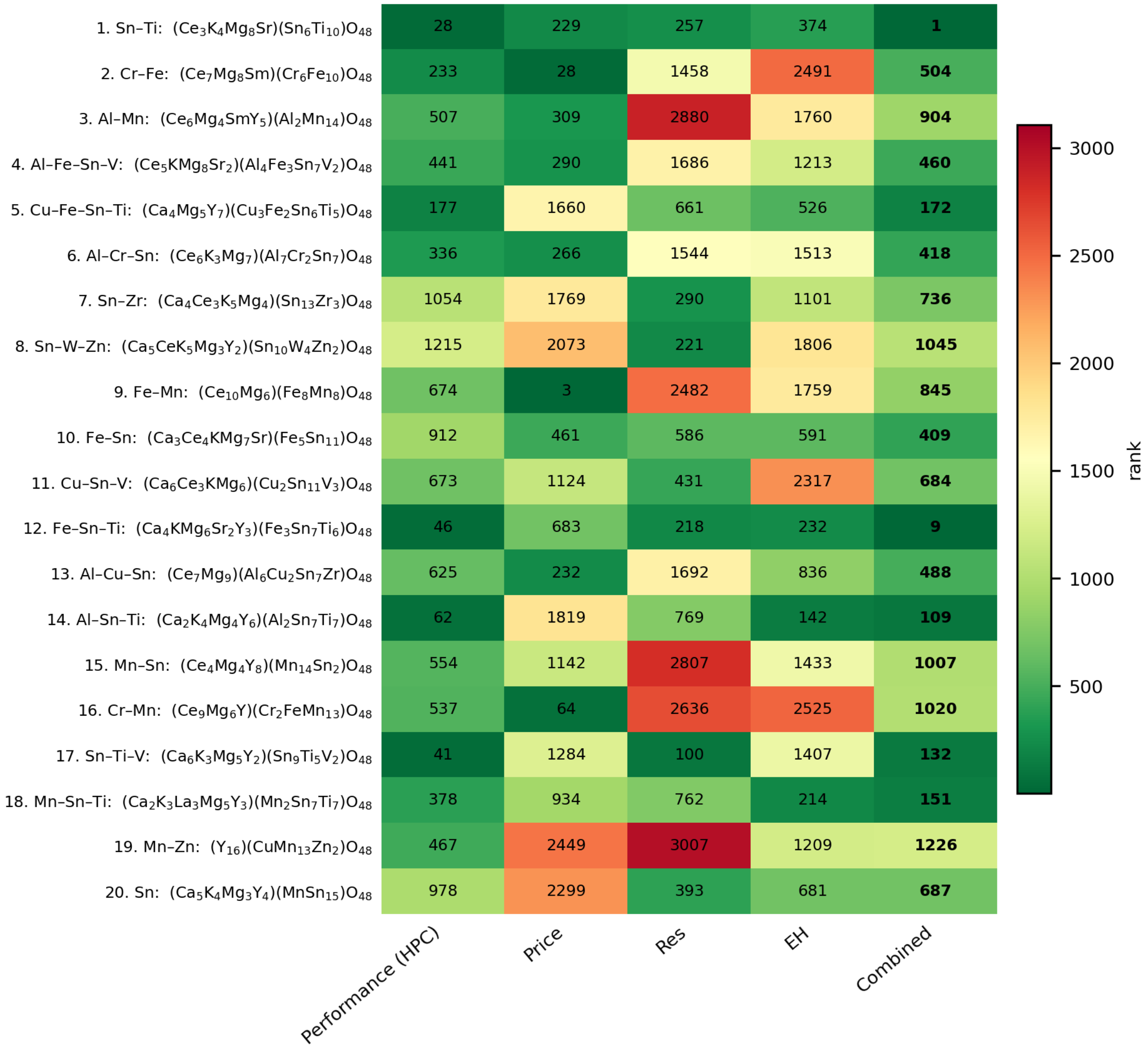

*Figure 7. Multi-objective ranking of 20 chemically diverse candidates, chosen by maximal marginal relevance (MMR) from the compounds of the starting set and the seven AL runs with HPC ≥ 1.0 mmol $H_2$ $g^{-1}$. Candidates are listed in the order in which they were picked: each pick balances a high combined score against a B-site chemistry that differs from those of the candidates already picked. Each row gives the candidate's B-site chemistry and its composition per 80-atom cell, $(A)_{16}(B)_{16}O_{48}$. Cells give the rank of the candidate (1 = best, out of 3108 compounds) for performance (HPC), raw-material price, supply resilience (Res), environmental and health (EH) impact, and the combined score (weights 0.6 / 0.1 / 0.15 and 0.15 for HPC / price / Res and EH, respectively). The color scale runs from green (high rank) to red (low rank).*

The HPC of all 20 candidates (Figure 8) lies between 2.22 and 2.84 mmol $H_2$ $g^{-1}$ (median 2.57), and six of them rank among the 250 best of the 3108 compounds with HPC > 0 on the combined score, including the overall best compound, $(Ce_3K_4Mg_8Sr)(Sn_6Ti_{10})O_{48}$. On the secondary criteria, expressed as percentiles among the compounds with HPC > 0, the candidates are mostly better than the median compound: 14 of the 20 for price, 13 for Res and 14 for EH, with median percentiles of 74, 75 and 61. They are not uniformly strong, and diversity comes at some cost, mainly for the transition-metal-rich picks. The five Mn-rich candidates lie between the 3$^{rd}$ and 20$^{th}$ percentile for supply resilience, and the Cr–Mn, Cr–Fe and Cu–Sn–V candidates at the 19$^{th}$, 20$^{th}$ and 26$^{th}$ percentile for EH impact. Sn remains the most common B-site element, appearing in 15 of the 20 picks, but it is combined with a wide range of metals (Ti, Zr, Al, W/Zn, V, Cu, Cr, Fe and Mn). Five candidates contain no Sn: Cr–Fe $(Ce_7Mg_8Sm)(Cr_6Fe_{10})O_{48}$, Fe–Mn $(Ce_{10}Mg_6)(Fe_8Mn_8)O_{48}$, Cr–Mn $(Ce_9Mg_6Y)(Cr_2FeMn_{13})O_{48}$, Al–Mn $(Ce_6Mg_4SmY_5)(Al_2Mn_{14})O_{48}$ and Mn–Zn $(Y_{16})(CuMn_{13}Zn_2)O_{48}$. The three cheapest candidates, Fe–Mn, Cr–Fe and Cr–Mn (1.6–2.3 USD $kg^{-1}$, all at or above the 98$^{th}$ percentile), are Sn-free and are cheap at the expense of Res,

EH or both. Only one candidate, $(Ce_3K_4Mg_8Sr)(Sn_6Ti_{10})O_{48}$, lies above the 85th percentile on all three secondary criteria. The picks differ in how many compounds they represent. The Sn–Ti (683 compounds), Al–Cr–Sn (228), Al–Sn–Ti (211), Sn–Ti–V (202), Sn (173), Cu–Sn–V (134) and Al–Mn (106) chemistries are densely sampled, whereas several others, e.g. Mn–Zn (1), Cr–Fe (4), Cr–Mn (12) and Mn–Sn–Ti (15), are represented by only a few compounds.

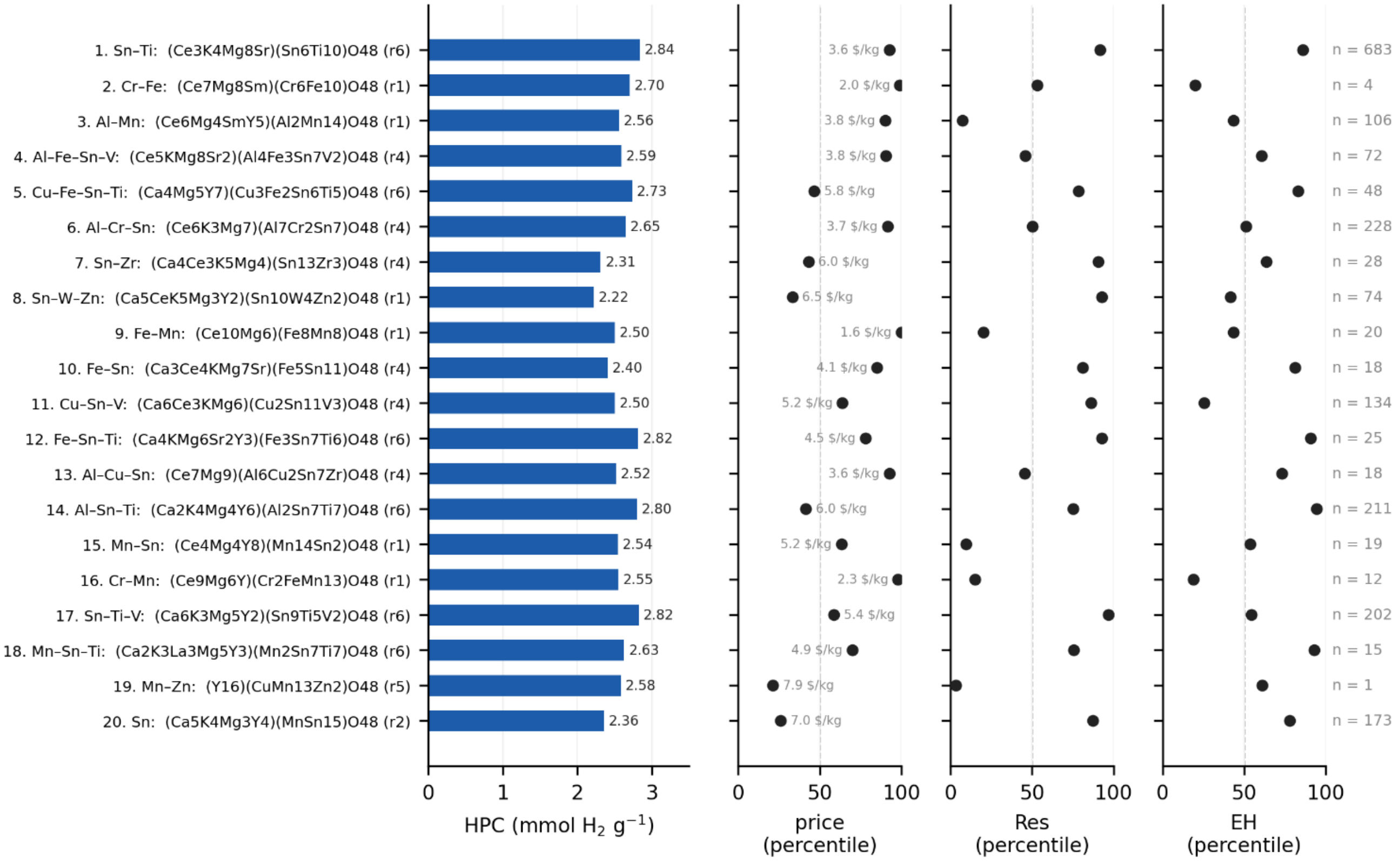

*Figure 8. 20 chemically diverse candidates drawn from the compounds with HPC >= 1.0 mmol $H_2 g^{-1}$ (2091 of the 3108 compounds with HPC > 0 from the starting set and the seven active-learning runs, cycles 1-50). Candidates are listed in pick order; labels give the pick number, the B-site chemistry, the composition as $A_{16}B_{16}O_{48}$ and the run in which the compound was found (r1–r7). n is the number of compounds assigned to the pick, i.e. whose B-site set is most similar to that of the pick. From the left: HPC value, raw-material price (including absolute value in USD/kg), supply resilience (Res) and environmental and health (EH) impact, each shown as the percentile among the 3108 compounds (100 = lower price, more resilient or less hazardous than others; dashed line = median).*

## 3.5. Comparison with experimentally reported hydrogen production capacities

To test the model against experiments, the HPC was calculated for seven perovskite oxygen carriers with experimentally reported hydrogen production capacities, using the reaction conditions of each study (Table 3). The FR and SR temperatures were taken directly from each study. Where a study reported gas conversions, these were converted into equivalent gas ratios. Where a ratio was not reported, the model was evaluated over a representative range, and the HPC is given as the resulting range. All compositions were modeled in an 80-atom supercell, so dopant fractions are approximated to the nearest value the cell allows.

For $La_{0.7}Sr_{0.3}FeO_3$ the model predicts 0.39–0.78 mmol $H_2$ $g^{-1}$ against 0.82–0.90 mmol $H_2$ $g^{-1}$ measured at 850 °C. For $La_{0.6}Sr_{0.4}FeO_3$ it predicts 0.25–0.38 mmol $H_2$ $g^{-1}$ against 0.21 mmol $H_2$ $g^{-1}$ measured at 900 °C. The model also reproduces the effect of Mo substitution in $SrFeO_3$ where $SrFe_{0.75}Mo_{0.25}O_3$ is predicted at 0.15–0.48 mmol $H_2$ $g^{-1}$, compared with no reversible capacity for undoped $SrFeO_3$, which matches the experimental ranking (~0.62 vs ~0.19 mmol $H_2$ $g^{-1}$). For the $La_{1-x}Ca_xMnO_3$ series, the predicted HPC of $LaMnO_3$ (0.14–0.80 mmol $H_2$ $g^{-1}$) covers the measured value of 0.147 mmol $H_2$ $g^{-1}$. For $La_{0.8}Ca_{0.2}MnO_3$, the measured 0.205 mmol $H_2$ $g^{-1}$ also lies within the predicted range, but this range (0–1.45 mmol $H_2$ $g^{-1}$) is too wide to be informative because the SR gas ratio was not reported and the HPC of the manganites is sensitive to it. The predicted ranges of $LaMnO_3$ and $La_{0.8}Ca_{0.2}MnO_3$ overlap, so the model does not resolve the higher measured HPC of the Ca-substituted material. The most direct comparison is $La_{0.5}Ca_{0.5}FeO_3$, which was tested in our own fluidized-bed experiments under the same reactor conditions as assumed in the active-learning runs (900 °C, $CO_2/CO$ = 0.11 in the FR and 800 °C, $H_2O/H_2$ = 4 in the SR) [71]. In this case, the gas ratios did not have to be estimated, and the model gives a single value, 0.13 mmol $H_2$ $g^{-1}$, against 0.15 mmol $H_2$ $g^{-1}$ measured. The model predicts the equilibrium, thermodynamic HPC of a single perovskite phase, while the experimental values also depend on reaction kinetics, how far the gas converts, the cycle length, and in many cases, the formation of secondary phases. Hence, absolute values are not expected to match, but rather the model should be tested for the right order of magnitude and for whether it ranks materials in the same order within one study. $La_{1-x}Sr_xFeO_3$, $LaMnO_3$ and $SrFe_{0.75}Mo_{0.25}O_3$ pass the order-of-magnitude test, and the model captures the effect of Mo substitution in $SrFeO_3$. The model underestimates the HPC of $La_{0.7}Sr_{0.3}FeO_3$ and $SrFe_{0.75}Mo_{0.25}O_3$, overestimates that of $La_{0.6}Sr_{0.4}FeO_3$ and comes closest for $La_{0.5}Ca_{0.5}FeO_3$, the only case in which the gas conditions were fully known. There is therefore no indication of a systematic bias, and much of the remaining spread reflects the unreported gas ratios.

The vacancy concentration for $SrFeO_3$ reaches the largest value the model considers ($\delta$ = 0.5) in both the FR and SR. As a result, no reversible oxygen swing is predicted in the model. Experimentally, $SrFeO_3$ and similar ferrites can reduce beyond $\delta$ = 0.5, partly through decomposition and phase changes. For example, $SrFeO_3$ reduces reversibly to SrO and Fe. Several of the higher capacities in Table 1, mainly from CLSMR studies, likely involve such decomposition. $La_{0.5}Sr_{0.5}FeO_3$ (3.34 mmol $H_2$ $g^{-1}$) [44], $CeO_2$/3DOM $LaFeO_3$ (3.38 mmol $H_2$ $g^{-1}$) [43] and $LaFe_{1-x}Co_xO3$ (≈ 4 mmol $H_2$ $g^{-1}$) [27] all exceed the 2.1–2.3 mmol $H_2$ $g^{-1}$ these compositions can deliver as a single perovskite phase at $\Delta\delta$ = 0.5. This indicates partial reduction to metallic Fe or Co. Such deep reduction raises the capacity, but free metallic Fe is prone to sintering and agglomeration, which can cause operational problems such as defluidization and loss of capacity over repeated cycles. When Fe instead remains in the perovskite lattice, the oxygen exchange takes place within a single crystal structure. The model, in contrast, is intended to design perovskites that split water without phase transition or decomposition, so such capacities are not directly comparable with the predicted HPC.

*Table 3. Hydrogen production capacities (HPC) of perovskite oxygen carriers reported in experimental studies, compared with the HPC calculated with the model under the experimental conditions.*

| Material<br><br>$ABO_3$ | Experimental gaseous conditions<br><br>FR, SR | Experimental HPC<br><br>[mmol $H_2$ $g^{-1}$] | Ref | Model FR conditions<br><br>$CO_2$/CO<br>T [°C ] | Model SR conditions<br><br>$H_2O$/H2<br>T [°C] | Modeled material<br><br>$A_{16}B_{16}O_{48}$ | Model HPC<br><br>mmol<br>$H_2$ $g^{-1}$ |
|---|---|---|---|---|---|---|---|
| $La_{0.7}Sr_{0.3}FeO_{3-\delta}$ | CO, $H_2O$ | 0.82–0.90 | [38] | 0.01-0.05<br>850 | 0.5-4<br>850 | $La_{11}Sr_5Fe_{16}O_{48}$ | 0.39-<br>0.78 |
| $La_{0.6}Sr_{0.4}FeO_{3-\delta}$ | CO, $H_2O$<br>55 % $H_2O$ conversion | 0.21 | [39] | 0.01-0.11<br>900 | 0.82<br>900 | $La_{10}Sr_6Fe_{16}O_{48}$ | 0.25-<br>0.38 |
| $SrFe_{0.75}Mo_{0.25}O_3$ | CO, $H_2O$ | ~0.62 | [40] | 0.01-0.11<br>700 | 0.5-4<br>700 | $Sr_{16}Fe_{12}Mo_4O_{48}$ | 0.15-<br>0.48 |
| $SrFeO_{3-\delta}$ | CO, $H_2O$ | ~0.19 | [40] | 0.01-0.11<br>700 | 0.5-4<br>700 | $Sr_{16}Fe_{16}O_{48}$ | 0* |
| $LaMnO_{3-\delta}$ | $CH_4$, $H_2O$ | 0.147 | [41] | 0.01-0.11<br>1000 | 0.59-13.3<br>1000 | $La_{16}Mn_{16}O_{48}$ | 0.14-<br>0.8 |
| $La_{0.8}Ca_{0.2}MnO_{3-\delta}$ | $CH_4$, $H_2O$ | 0.205 | [41] | 0.01-0.11<br>1000 | 0.59-13.3<br>1000 | $La_{13}Ca_3Mn_{16}O_{48}$ | 0-1.45 |
| $La_{0.5}Ca_{0.5}FeO_{3-\delta}$ | CO+$H_2$, $H_2O$ | 0.15 | [71] | 0.11<br>900 | 4<br>800 | $La_8Ca_8Fe_{16}O_{48}$ | 0.13 |

* The vacancy concentration reaches the limit of δ = 0.5 under both FR and SR conditions, so no reversible oxygen exchange is captured within the model.

The predicted HPC of the 20 candidates, 2.22–2.84 mmol $H_2$ $g^{-1}$, are all above 1.99 mmol $H_2$ $g^{-1}$ of iron oxides [7, 34]. Iron oxides, however, reach this capacity through a phase transition between wüstite and magnetite, whereas the perovskite candidates considered here are intended to exchange oxygen within a single crystal structure. Beyond these comparisons, the candidates can be related to the simple perovskites from which their B-site chemistries derive. $LaMnO_{3-\delta}$, $CaMnO_{3-\delta}$, $LaFeO_{3-\delta}$, $CaFeO_{3-\delta}$, and $YFeO_{3-\delta}$ are all established oxygen carriers. $CaMnO_{3-\delta}$ has been studied extensively for chemical looping with oxygen uncoupling [72], $LaFeO_{3-\delta}$ and $LaMnO_{3-\delta}$ for chemical looping hydrogen production [73, 74], and $CaFeO_{3-\delta}$, known to form brownmillerite $Ca_2Fe_2O_5$, for chemical looping hydrogen and syngas production [75]. $CaSnO_{3-\delta}$ has been scarcely investigated for chemical looping combustion. The perovskite structure stabilized the low-melting Sn oxide over repeated cycles but showed a slower reduction than other carriers, which was improved by partial Co-substitution on the A-site. [76] $YFeO_{3-\delta}$ and Sr-substituted $Sr_xY_{1-x}FeO_{3-\delta}$ (x = 0–0.3) have been studied for chemical looping oxidative dehydrogenation (CL-ODH) of ethane, where Sr substitution increased the oxygen vacancy concentration and moderated the lattice oxygen reactivity, while pristine $YFeO_{3-\delta}$ was more reactive, which is undesirable for CL-ODH but favorable for full fuel conversion. [77] Under the conditions of the active-learning search, the model gives these perovskites an HPC of at most 0.76 mmol $H_2$ $g^{-1}$ ($LaMnO_{3-\delta}$ 0.47, $LaFeO_{3-\delta}$ 0.53 and $YFeO_{3-\delta}$ 0.76 mmol $H_2$ $g^{-1}$) and 0 for $CaFeO_{3-\delta}$ and $CaSnO_{3-\delta}$, compared with 2.99 mmol $H_2$ $g^{-1}$ for the best-performing composition.

# 4. Conclusion

This study demonstrates how an autonomous active-learning framework can accelerate the discovery of perovskite oxygen carriers for chemical looping hydrogen production. By combining uncertainty-aware exploration with thermodynamic constraints representative of the chemical looping process, the framework navigated a large compositional design space and raised the mean HPC of the acquired compositions from 0.41 mmol $H_2$ $g^{-1}$ in the starting set to 1.56 mmol $H_2$ $g^{-1}$ in the first cycle and 2.05 mmol $H_2$ $g^{-1}$ in cycle 50. All seven independent runs showed the same increase, although their final levels differed.

All seven runs enriched Sn on the B-site mostly together with Mg, Ce, K and Y. In most runs the best compositions combined Sn with Ti, whereas one also explored an Mn-rich domain with a lower average HPC. The model was also benchmarked against perovskite oxygen carriers characterized in the literature. It reproduced the measured HPC of $La_{1-x}Sr_xFeO_3$, $LaMnO_3$ and Mo-substituted $SrFeO_3$ in order of magnitude and captured the increase in HPC from Mo substitution in SrFeO3. Because the model gives the equilibrium HPC of a single perovskite phase, it is intended to screen and rank candidates rather than to predict absolute capacities. For $La_{0.5}Ca_{0.5}FeO_3$, tested in a batch fluidized-bed reactor, the predicted HPC (0.13 mmol $H_2$ $g^{-1}$) was close to the measured value (0.15 mmol $H_2$ $g^{-1}$).

More importantly, whereas the AL search was driven by acquiring high-performing compositions, the final candidates were not selected on performance only. Incorporating raw material price, supply resilience, and potential environmental and health impacts associated with precursor handling introduced practical considerations relevant to large-scale deployment. Based on the multi-objective ranking, 20 chemically diverse candidates are suggested reaching 2.22–2.84 mmol $H_2$ $g^{-1}$ (median 2.57). This is above the 1.99 mmol $H_2$ $g^{-1}$ of iron oxide, and three to four times the highest HPC the model predicts for the simple bimetallic perovskites. However, performance and practicality do not always coincide e.g. the three cheapest candidates, Fe–Mn, Cr–Fe and Cr–Mn compositions, are cheap at the expense of supply resilience, environmental and health impact or both, and all five Mn-rich candidates lie at or below the 20$^{th}$ percentile for supply resilience. Sn–Ti-based compositions offered the best balance and more specifically $(Ce_3K_4Mg_8Sr)(Sn_6Ti_{10})O_{48}$ has the best combined rank and is the only candidate above the 85$^{th}$ percentile on all three secondary criteria, followed by the Fe–Sn–Ti candidate $(Ca_4KMg_6Sr_2Y_3)(Fe_3Sn_7Ti_6)O_{48}$. This multi-objective approach provides a means of balancing material performance with resource availability and precursor-related considerations, thereby moving materials discovery beyond performance-based screening toward process-oriented candidate selection.

Overall, the results demonstrate the potential of combining autonomous materials discovery, thermodynamic screening, multi-objective candidate selection, and benchmarking against experimental data to identify promising oxygen carrier compositions for chemical looping hydrogen production and other chemical looping processes. Future work should synthesize the highest-ranked candidates, in particular the Sn–Ti and Fe–Sn–Ti compositions, and evaluate their phase stability, redox kinetics, cyclic stability and hydrogen production under realistic chemical looping conditions.

# Data availability

The code used for the active-learning workflow, HPC calculations and candidate selection is openly available at https://github.com/stanicic-chalmers/DesigningPerovskitesForCLHP and archived on Zenodo (release v1.0.0, https://doi.org/10.5281/zenodo.24393271). This archive also contains the databases of the seven active-learning runs (r1–r7), including the starting dataset, the configuration files used for each run, and the scripts and data used to produce the figures in this paper. The SQS structures and phonon data of the starting dataset are available on Zenodo at https://doi.org/10.5281/zenodo.24197403.

## Acknowledgements

This work was supported by the Wallenberg Initiative Materials Science for Sustainability (WISE) funded by the Knut and Alice Wallenberg Foundation. Computational resources were provided on the Berzelius system (Berzelius-2026-135 and Berzelius-2026-298), funded by the Knut and Alice Wallenberg Foundation and operated by the National Academic Infrastructure for Supercomputing in Sweden (NAISS). NAISS also provided computational resources, partially funded by the Swedish Research Council through grant agreement no. NAISS 2025/5-313, NAISS 2025/5-530, NAISS 2026/3-535, NAISS 2026/3-498, and NAISS 2026/3-291.

## References

[1] I. Kountouris, R. Bramstoft, T. Madsen, J. Gea-Bermúdez, M. Münster, and D. Keles, "A unified European hydrogen infrastructure planning to support the rapid scale-up of hydrogen production," *Nature Communications,* vol. 15, no. 1, p. 5517, 2024, doi: 10.1038/s41467-024-49867-w.

[2] International Energy Agency, "Hydrogen Patents for a Clean Energy Future," 2023. [Online]. Available: https://www.iea.org/reports/hydrogen-patents-for-a-clean-energy-future (accessed 2025).

[3] A. G. Al-Gamal, F. Yehia, M. R. Elmasry, M. A. Abo El-Khair, H. S. Kandeel, A. M. Elseman, D.-H. Kim, and K. I. Kabel, "Perovskite materials for hydrogen evolution: Processes, challenges and future perspectives," *International Journal of Hydrogen Energy,* vol. 79, pp. 1113-1138, 2024, doi: 10.1016/j.ijhydene.2024.07.039.

[4] R. J. Carrillo and J. R. Scheffe, "Advances and trends in redox materials for solar thermochemical fuel production," *Solar Energy,* vol. 156, pp. 3-20, 2017, doi: 10.1016/j.solener.2017.05.032.

[5] L. Zeng, Z. Cheng, J. A. Fan, L.-S. Fan, and J. Gong, "Metal oxide redox chemistry for chemical looping processes," *Nature Reviews Chemistry,* vol. 2, no. 11, pp. 349-364, 2018, doi: 10.1038/s41570-018-0046-2.

[6] W. Chang, Y. Hu, W. Xu, C. Huang, H. Chen, J. He, Y. Han, Y. Zhu, X. Ma, and X. Wang, "Recent Advances of Oxygen Carriers for Hydrogen Production via Chemical Looping Water-Splitting," *Catalysts,* vol. 13, no. 2, p. 279, 2023, doi: 10.3390/catal13020279.

[7] I. Staničić, R. Faust, P. Knutsson, V. Purnomo, H. Leion, T. Mattisson, and M. Rydén, "Performance of magnetite fines for continuous hydrogen production," *International Journal of Hydrogen Energy,* vol. 139, pp. 753-765, 2025, doi: 10.1016/j.ijhydene.2025.05.223.

[8] A. A. Emery, J. E. Saal, S. Kirklin, V. I. Hegde, and C. Wolverton, "High-Throughput Computational Screening of Perovskites for Thermochemical Water Splitting Applications," *Chemistry of Materials,* vol. 28, no. 16, pp. 5621-5634, 2016, doi: 10.1021/acs.chemmater.6b01182.

[9] T. Lookman, P. V. Balachandran, D. Xue, and R. Yuan, "Active learning in materials science with emphasis on adaptive sampling using uncertainties for targeted design," *npj Computational Materials,* vol. 5, no. 1, p. 21, 2019, doi: 10.1038/s41524-019-0153-8.

[10] X. Wang, Y. Gao, E. Krzystowczyk, S. Iftikhar, J. Dou, R. Cai, H. Wang, C. Ruan, S. Ye, and F. Li, "High-throughput oxygen chemical potential engineering of perovskite oxides for chemical looping applications," *Energy & Environmental Science, vol. 15*, no. 4, pp. 1512-1528, 2022, doi: 10.1039/D1EE02889H.

[11] A. H. McDaniel, E. C. Miller, D. Arifin, A. Ambrosini, E. N. Coker, R. O'Hayre, W. C. Chueh, and J. Tong, "Sr- and Mn-doped LaAlO3−δ for solar thermochemical H2 and CO production," *Energy & Environmental Science,* vol. 6, no. 8, pp. 2424-2428, 2013, doi: 10.1039/c3ee41372a.

[12] C.-K. Yang, Y. Yamazaki, A. Aydin, and S. M. Haile, "Thermodynamic and kinetic assessments of strontium-doped lanthanum manganite perovskites for two-step thermochemical water splitting," *Journal of Materials Chemistry A,* vol. 2, no. 33, pp. 13612-13623, 2014, doi: 10.1039/c4ta02694b.

[13] S. Dey, B. S. Naidu, and C. N. R. Rao, "Ln0.5A0.5MnO3 (Ln=Lanthanide, A= Ca, Sr) Perovskites Exhibiting Remarkable Performance in the Thermochemical Generation of CO and H2 from CO2 and H2O," *Chemistry – A European Journal,* vol. 21, no. 19, pp. 7077-7081, 2015, doi: 10.1002/chem.201500442.

[14] T. Cooper, J. R. Scheffe, M. E. Galvez, R. Jacot, G. Patzke, and A. Steinfeld, "Lanthanum Manganite Perovskites with Ca/Sr A-site and Al B-site Doping as Effective Oxygen Exchange Materials for Solar Thermochemical Fuel Production," *Energy Technology,* vol. 3, no. 11, pp. 1130-1142, 2015, doi: 10.1002/ente.201500226.

[15] D. R. Barcellos, M. D. Sanders, J. Tong, A. H. McDaniel, and R. P. O'Hayre, "$BaCe_{0.25}Mn_{0.75}O_{3-\delta}$—a promising perovskite-type oxide for solar thermochemical hydrogen production," *Energy & Environmental Science,* vol. 11, no. 11, pp. 3256-3265, 2018, doi: 10.1039/c8ee01989d.

[16] X. Qian, J. He, E. Mastronardo, B. Baldassarri, C. Wolverton, and S. M. Haile, "Favorable Redox Thermodynamics of $SrTi_{0.5}Mn_{0.5}O_{3-\delta}$ in Solar Thermochemical Water Splitting," *Chemistry of Materials,* vol. 32, no. 21, pp. 9335-9346, 2020, doi: 10.1021/acs.chemmater.0c03278.

[17] X. Qian, J. He, E. Mastronardo, B. Baldassarri, W. Yuan, C. Wolverton, and S. M. Haile, "Outstanding Properties and Performance of $CaTi_{0.5}Mn_{0.5}O_{3-\delta}$ for Solar-Driven Thermochemical Hydrogen Production," *Matter,* vol. 4, no. 2, pp. 688-708, 2021, doi: 10.1016/j.matt.2020.11.016.

[18] S. J. Heo, M. Sanders, R. O'Hayre, and A. Zakutayev, "Double-Site Substitution of Ce into (Ba, Sr)$MnO_3$ Perovskites for Solar Thermochemical Hydrogen Production," *ACS Energy Letters,* vol. 6, no. 9, pp. 3037-3043, 2021, doi: 10.1021/acsenergylett.1c01214.

[19] A. M. Bergeson-Keller, M. D. Sanders, and R. P. O'Hayre, "Reduction Thermodynamics of $Sr_{1-x}Ce_xMnO_3$ and $Ce_xSr_{2-x}MnO_4$ Perovskites for Solar Thermochemical Hydrogen Production," *Energy Technology,* vol. 10, no. 1, p. 2100515, 2022, doi: 10.1002/ente.202100515.

[20] J. M. Naik, B. Bulfin, C. A. Triana, D. C. Stoian, and G. R. Patzke, "Cation-Deficient Ce-Substituted Perovskite Oxides with Dual-Redox Active Sites for Thermochemical Applications," *ACS Applied Materials & Interfaces,* vol. 15, no. 1, pp. 806-817, 2023, doi: 10.1021/acsami.2c15169.

[21] C. Liu, D. Zhang, W. Li, J. A. Trindell, K. A. King, S. R. Bishop, J. D. Sugar, A. H. McDaniel, A. I. Smith, P. A. Salinas, E. N. Coker, A. L. Clauser, M. Velayutham, J. C. Neuefeind, J. Yang, H. A. De Santiago, L. Ma, Y. Wang, Q. Wang, W. Li, Q. Wang, Q. Li, H. Tian, H. N. N. Tran, X. Li, B. Robinson, A. M. Deibel, G. Collins, N. A. Thieu, J. Hu, V. V. Khramtsov, J. Luo, and X. Liu, "Manganese-based A-site high-entropy perovskite oxide for solar thermochemical hydrogen production," *Journal of Materials Chemistry A,* vol. 12, no. 7, pp. 3910-3922, 2024, doi: 10.1039/d3ta03554a.

[22] D. Zhang, H. A. De Santiago, B. Xu, C. Liu, J. A. Trindell, W. Li, J. Park, M. A. Rodriguez, E. N. Coker, J. D. Sugar, A. H. McDaniel, S. Lany, L. Ma, Y. Wang, G. Collins, H. Tian, W. Li, Y. Qi, X. Liu, and J. Luo, "Compositionally Complex Perovskite Oxides for Solar Thermochemical Water Splitting," *Chemistry of Materials,* vol. 35, no. 5, pp. 1901-1915, 2023, doi: 10.1021/acs.chemmater.2c03054.

[23] A. Le Gal, M. Vallès, A. Julbe, and S. Abanades, "Thermochemical Properties of High Entropy Oxides Used as Redox-Active Materials in Two-Step Solar Fuel Production Cycles," *Catalysts,* vol. 12, no. 10, p. 1116, 2022, doi: 10.3390/catal12101116.

[24] F. He and F. Li, "Perovskite promoted iron oxide for hybrid water-splitting and syngas generation with exceptional conversion," *Energy & Environmental Science,* vol. 8, no. 2, pp. 535-539, 2015, doi: 10.1039/c4ee03431g.

[25] V. P. Haribal, F. He, A. Mishra, and F. Li, "Iron-Doped $BaMnO_3$ for Hybrid Water Splitting and Syngas Generation," *ChemSusChem,* vol. 10, no. 17, pp. 3402-3408, 2017, doi: 10.1002/cssc.201700699.

[26] C. Dueso, C. Thompson, and I. Metcalfe, "High-stability, high-capacity oxygen carriers: Iron oxide-perovskite composite materials for hydrogen production by chemical looping," *Applied Energy,* vol. 157, pp. 382-390, 2015, doi: 10.1016/j.apenergy.2015.05.062.

[27] K. Zhao, F. He, Z. Huang, G. Wei, A. Zheng, H. Li, and Z. Zhao, "Perovskite-type oxides $LaFe_{1-x}Co_xO_3$ for chemical looping steam methane reforming to syngas and hydrogen co-production," *Applied Energy,* vol. 168, pp. 193-203, 2016, doi: 10.1016/j.apenergy.2016.01.052.

[28] K. Zhao, L. Li, A. Zheng, Z. Huang, F. He, Y. Shen, G. Wei, H. Li, and Z. Zhao, "Synergistic improvements in stability and performance of the double perovskite-type oxides $La_{2-x}Sr_xFeCoO_6$ for chemical looping steam methane reforming," *Applied Energy,* vol. 197, pp. 393-404, 2017, doi: 10.1016/j.apenergy.2017.04.049.

[29] J. Ha, H. S. Kim, H. Kim, Y. Kim, S. A. N. Asih, and J. W. Lee, "Modulating Oxygen Transfer via A-Site Doping in $LaFeO_3$ for Coke-Resistant Chemical Looping Steam Methane Reforming," *ChemSusChem,* vol. 18, no. 24, p. e202501667, 2025, doi: 10.1002/cssc.202501667.

[30] K. Zhao, R. Zhang, Y. Gao, Y. Lin, A. Liu, X. Wang, A. Zheng, Z. Huang, and Z. Zhao, "High syngas selectivity and near pure hydrogen production in perovskite oxygen carriers for chemical looping steam methane reforming," *Fuel Processing Technology,* vol. 236, p. 107398, 2022, doi: 10.1016/j.fuproc.2022.107398.

[31] S. Abanades, X. Wang, and S. Chuayboon, "Chemical Looping $CH_4$ Reforming Through Isothermal Two-Step Redox Cycling of $SrFeO_3$ Oxygen Carrier in a Tubular Solar Reactor," *Molecules,* vol. 30, no. 5, p. 1076, 2025, doi: 10.3390/molecules30051076.

[32] R. B. Wexler, G. Sai Gautam, R. T. Bell, S. Shulda, N. A. Strange, J. A. Trindell, J. D. Sugar, E. Nygren, S. Sainio, A. H. McDaniel, D. Ginley, E. A. Carter, and E. B. Stechel, "Multiple and nonlocal cation redox in Ca–Ce–Ti–Mn oxide perovskites for solar thermochemical applications," *Energy & Environmental Science,* vol. 16, no. 6, pp. 2550-2560, 2023, doi: 10.1039/d3ee00234a.

[33] E. Gager, M. Frye, D. McCord, J. Scheffe, and J. C. Nino, "Reticulated porous lanthanum strontium manganite structures for solar thermochemical hydrogen production," *International Journal of Hydrogen Energy,* vol. 47, no. 73, pp. 31152-31164, 2022, doi: 10.1016/j.ijhydene.2022.07.052.

[34] I. Staničić, T. Mattisson, and M. Rydén, "Apparent Kinetics of the Reduction of Magnetite Fines with Methane in Fluidized Beds," presented at the 25th International Conference on Fluidized Bed Conversion, Nanjing, China, 2025.

[35] D. Hosseini, F. Donat, P. M. Abdala, S. M. Kim, A. M. Kierzkowska, and C. R. Müller, "Reversible Exsolution of Dopant Improves the Performance of $Ca_2Fe_2O_5$ for Chemical Looping Hydrogen Production," *ACS Applied Materials & Interfaces,* vol. 11, no. 20, pp. 18276-18284, 2019, doi: 10.1021/acsami.8b16732.

[36] Y. Long, K. Yang, Z. Gu, S. Lin, D. Li, X. Zhu, H. Wang, and K. Li, "Hydrogen generation from water splitting over polyfunctional perovskite oxygen carriers by using coke oven gas as reducing agent," *Applied Catalysis B: Environmental,* vol. 301, p. 120778, 2022, doi: 10.1016/j.apcatb.2021.120778.

[37] P. V. Ambeth, M. H. Ahmed, and M. Konarova, "Enhanced Oxygen Evolution in Perovskite Materials for Syngas Production and Water Splitting in Chemical Looping," *ChemCatChem,* vol. 18, no. 2, p. e202401980, 2026, doi: 10.1002/cctc.202401980.

[38] A. Murugan, A. Thursfield, and I. S. Metcalfe, "A chemical looping process for hydrogen production using iron-containing perovskites," *Energy & Environmental Science,* vol. 4, no. 11, pp. 4639-4649, 2011, doi: 10.1039/c1ee02142g.

[39] A. Zaidi, C. de Leeuwe, S. Chansai, C. Hardacre, A. Garforth, C. Parlett, and V. Spallina, "Development of iron-nickel containing perovskites with increased oxygen carrier capacity for chemical looping $H_2$ production," *Journal of Environmental Chemical Engineering,* vol. 13, no. 1, p. 115069, 2025, doi: 10.1016/j.jece.2024.115069.

[40] A. J. Carrillo, D. Catalán-Martínez, S. Domínguez, M. Balaguer, R. J. Meyer, K. Wang, and J. M. Serra, "$Sr_2Fe_{1.5}Mo_{0.5}O_{6-\delta}$ double perovskite for chemical-looping hydrogen production from water splitting," *Chemical Engineering Journal,* vol. 544, p. 178254, 2026, doi: 10.1016/j.cej.2026.178254.

[41] M. Moschos, A. Evdou, and V. Zaspalis, "$(La_{1-x}Ca_x)MnO_{3-\delta}$ (x = 0, 0.2, 0.3, 0.4) Perovskites as Redox Catalysts in Chemical Looping Hydrogen Production Process: The Relation between Defect Chemistry and Redox Performance," *Catalysts,* vol. 14, no. 7, p. 431, 2024, doi: 10.3390/catal14070431.

[42] K. Zhao, F. He, Z. Huang, A. Zheng, H. Li, and Z. Zhao, "Three-dimensionally ordered macroporous LaFeO3 perovskites for chemical-looping steam reforming of methane," *International Journal of Hydrogen Energy,* vol. 39, no. 7, pp. 3243-3252, 2014, doi: 10.1016/j.ijhydene.2013.12.046.

[43] Y. Zheng, K. Li, H. Wang, D. Tian, Y. Wang, X. Zhu, Y. Wei, M. Zheng, and Y. Luo, "Designed oxygen carriers from macroporous LaFeO3 supported CeO2 for chemical-looping reforming of methane," *Applied Catalysis B: Environmental,* vol. 202, pp. 51-63, 2017, doi: 10.1016/j.apcatb.2016.08.024.

[44] K. Yang, Z. Gu, Y. Long, S. Lin, C. Lu, X. Zhu, H. Wang, and K. Li, "Hydrogen production via chemical looping reforming of coke oven gas," *Green Energy & Environment,* vol. 6, no. 5, pp. 678-692, 2021, doi: 10.1016/j.gee.2020.06.027.

[45] X. Yin, S. Wang, B. Wang, and L. Shen, "Chemical looping steam methane reforming using Al doped LaMnO3+δ perovskites as oxygen carriers," *International Journal of Hydrogen Energy,* vol. 46, no. 67, pp. 33375-33387, 2021, doi: 10.1016/j.ijhydene.2021.07.192.

[46] J. Vieten, B. Bulfin, P. Huck, M. Horton, D. Guban, L. Zhu, Y. Lu, K. A. Persson, M. Roeb, and C. Sattler, "Materials design of perovskite solid solutions for thermochemical applications," *Energy & Environmental Science,* vol. 12, no. 4, pp. 1369-1384, 2019, doi: 10.1039/c9ee00085b.

[47] G. Sai Gautam, E. B. Stechel, and E. A. Carter, "Exploring Ca–Ce–M–O (M = 3d Transition Metal) Oxide Perovskites for Solar Thermochemical Applications," *Chemistry of Materials,* vol. 32, no. 23, pp. 9964-9982, 2020, doi: 10.1021/acs.chemmater.0c02912.

[48] R. B. Wexler, G. S. Gautam, E. B. Stechel, and E. A. Carter, "Factors Governing Oxygen Vacancy Formation in Oxide Perovskites," *Journal of the American Chemical Society,* vol. 143, no. 33, pp. 13212-13227, 2021, doi: 10.1021/jacs.1c05570.

[49] J. Brorsson, I. Staničić, J. Gastaldi, T. Mattisson, and A. Hellman, "Thermodynamic properties for metal oxides from first-principles," *Computational Materials Science*, vol. 233, p. 112690, 2024, doi: 10.1016/j.commatsci.2023.112690.

[50] J. Perry, L. Molina, A. de la Calle, R. Peño, T. W. Jones, M. V. Ganduglia-Pirovano, S. Jiménez-Fernández, S. W. Donne, J. M. Coronado, and A. Bayon, "Discovery of materials for solar thermochemical hydrogen combining machine learning, computational chemistry, experiments and system simulations," *npj Computational Materials,* vol. 11, no. 1, p. 247, 2025, doi: 10.1038/s41524-025-01726-y.

[51] J. Brorsson, H. Klein Moberg, J. Hildingsson, J. Gastaldi, T. Mattisson, and A. Hellman, "Data-Efficient Design of High-Entropy Oxygen Carriers for Chemical Looping Using Active Learning," *ACS Materials Au,* vol. 6, no. 2, pp. 319-326, 2026, doi: 10.1021/acsmaterialsau.5c00230.

[52] J. Brorsson, H. Klein Moberg, J. Gastaldi, J. Hildingsson, T. Mattisson, and A. Hellman, "Discovery of high-entropy perovskite oxygen carriers for chemical looping applications via an autonomous active learning protocol," *Materials Today Energy,* vol. 57, p. 102239, 2026, doi: 10.1016/j.mtener.2026.102239.

[53] C. J. Bartel, C. Sutton, B. R. Goldsmith, R. Ouyang, C. B. Musgrave, L. M. Ghiringhelli, and M. Scheffler, "New tolerance factor to predict the stability of perovskite oxides and halides," *Science Advances,* vol. 5, no. 2, p. eaav0693, 2019, doi: 10.1126/sciadv.aav0693.

[54] A. Zunger, S. H. Wei, L. G. Ferreira, and J. E. Bernard, "Special quasirandom structures," *Physical Review Letters,* vol. 65, no. 3, pp. 353-356, 1990, doi: 10.1103/PhysRevLett.65.353.

[55] M. Ångqvist, W. A. Muñoz, J. M. Rahm, E. Fransson, C. Durniak, P. Rozyczko, T. H. Rod, and P. Erhart, "ICET – A Python Library for Constructing and Sampling Alloy Cluster Expansions," *Advanced Theory and Simulations,* vol. 2, no. 7, p. 1900015, 2019, doi: 10.1002/adts.201900015.

[56] B. Deng, P. Zhong, K. Jun, J. Riebesell, K. Han, C. J. Bartel, and G. Ceder, "CHGNet as a pretrained universal neural network potential for charge-informed atomistic modelling," *Nature Machine Intelligence,* vol. 5, no. 9, pp. 1031-1041, 2023, doi: 10.1038/s42256-023-00716-3.

[57] A. Hjorth Larsen, J. J. Mortensen, J. Blomqvist, I. E. Castelli, R. Christensen, M. Dułak, J. Friis, M. N. Groves, B. Hammer, C. Hargus, et al., "The atomic simulation environment—a Python library for working with atoms," *Journal of Physics: Condensed Matter,* vol. 29, no. 27, p. 273002, 2017, doi: 10.1088/1361-648X/aa680e.

[58] A. Togo, L. Chaput, T. Tadano, and I. Tanaka, "Implementation strategies in phonopy and phono3py," *Journal of Physics: Condensed Matter,* vol. 35, no. 35, p. 353001, 2023, doi: 10.1088/1361-648X/acd831.

[59] T. C. Allison, "NIST-JANAF Thermochemical Tables - SRD 13," National Institute of Standards and Technology, 2013, doi: 10.18434/T42S31.

[60] B. Lakshminarayanan, A. Pritzel, and C. Blundell, "Simple and scalable predictive uncertainty estimation using deep ensembles," in *Advances in Neural Information Processing Systems*, vol. 30, Long Beach, CA, USA, 2017.

[61] T. Pearce, F. Leibfried, and A. Brintrup, "Uncertainty in neural networks: Approximately Bayesian ensembling," in *Proceedings of the International Conference on Artificial Intelligence and Statistics (AISTATS), PMLR, vol. 108, 2020, pp. 234-244*.

[62] X. Lu and B. Van Roy, "Ensemble sampling," *Advances in Neural Information Processing Systems,* vol. 30, 2017.

[63] I. Tolstikhin, O. Bousquet, S. Gelly, and B. Schoelkopf, "Wasserstein auto-encoders," *arXiv preprint arXiv:1711.01558,* 2017.

[64] L. McInnes, J. Healy, and J. Melville, "UMAP: Uniform manifold approximation and projection for dimension reduction," *arXiv preprint arXiv:1802.03426*, 2018.

[65] Leonland, "Chemical elements by market price." [Online]. Available: http://www.leonland.de/elements_by_price/en/list (accessed July 1, 2026).

[66] European Commission, Directorate-General for Internal Market, Industry, Entrepreneurship and SMEs, M. Grohol, and C. Veeh, *Study on the critical raw materials for the EU 2023 – Final report*. Publications Office of the European Union, 2023.

[67] European Parliament and Council of the European Union, "Regulation (EU) 2024/1252 (Critical Raw Materials Act)," Official Journal of the European Union, 2024. [Online]. Available: http://data.europa.eu/eli/reg/2024/1252/2024-05-03

[68] J. Carbonell and J. Goldstein, "The use of MMR, diversity-based reranking for reordering documents and producing summaries," in *Proceedings of the 21st Annual International ACM SIGIR Conference on Research and Development in Information Retrieval*, Melbourne, Australia, 1998, pp. 335-336, doi: 10.1145/290941.291025.

[69] Y. A. Mastrikov, D. Gryaznov, G. Zvejnieks, M. N. Sokolov, M. Putniņa, and E. A. Kotomin, "Sr Doping and Oxygen Vacancy Formation in La1−*x*Sr*x*ScO3−*δ* Solid Solutions: Computational Modelling," *Crystals*, vol. 12, no. 9, p. 1300, 2022, doi: 10.3390/cryst12091300.

[70] E. Bucher and W. Sitte, "Defect Chemical Modeling of (La,Sr)(Co,Fe)O3-δ," *Journal of Electroceramics*, vol. 13, pp. 779-784, 2004, doi: 10.1007/s10832-004-5192-x.

[71] A. Klute, "Perovskite Materials for Hydrogen Production - Synthesis and Evaluation of Perovskite Metal Oxides for Hydrogen Production Using Chemical Looping Water Splitting," M.S. thesis, Department of Environmental and Energy Sciences, Chalmers University of Technology, 2026.

[72] J. Gastaldi, J. Brorsson, I. Staničić, A. Hellman, and T. Mattisson, "First-Principles Estimation of Thermodynamic Properties and Phase Stability of CaMnO3−δ for Chemical-Looping Combustion," *Energy & Fuels,* vol. 39, no. 19, pp. 9113-9120, 2025, doi: 10.1021/acs.energyfuels.5c00267.

[73] A. Evdou and V. Zaspalis, "Comparative Evaluation of Various ABO3 Perovskites (A = La, Ca, Sr; B = Mn, Fe) as Oxygen Carrier Materials in Chemical Looping Hydrogen Production," *Hydrogen,* vol. 6, no. 2, p. 27, 2025, doi: 10.3390/hydrogen6020027.

[74] M. Orfila, M. Linares, R. Molina, J. Á. Botas, R. Sanz, and J. Marugán, "Perovskite materials for hydrogen production by thermochemical water splitting," *International Journal of Hydrogen Energy,* vol. 41, no. 42, pp. 19329-19338, 2016, doi: 10.1016/j.ijhydene.2016.07.041.

[75] C. Liu, W. Wang, and D. Chen, "Hydrogen-Rich Syngas Production from Chemical Looping Gasification of Biomass Char with CaMn1–xFexO3," *Energy & Fuels,* vol. 32, no. 9, pp. 9541-9550, 2018, doi: 10.1021/acs.energyfuels.8b01836.

[76] S. H. Baek, S. J. Park, H.-D. Kim, and J. M. Sohn, "A Study on the Redox Characteristics of Oxygen Carriers CaSnO3 and Materials Partially Substituted with Co and Ce for Chemical Looping Combustion," *Korean Journal of Chemical Engineering,* vol. 42, no. 4, pp. 791-801, 2025, doi: 10.1007/s11814-024-00337-5.

[77] S. J. Park, S. H. Baek, H. D. Kim, and J. M. Sohn, "Compositional engineering of Sr-substituted perovskite oxygen carriers for selective ethylene production via chemical looping oxidative dehydrogenation," *Fuel Processing Technology,* vol. 285, p. 108430, 2026, doi: 10.1016/j.fuproc.2026.108430.

# Supplementary material

## Designing Perovskites for Chemical Looping Hydrogen Production

Ivana Staničić[1,2*], Alfred Juhlin Onbeck[2], Anders Hellman[3], Tobias Mattisson[2], Magnus Rydén[1,2]

[1] Wallenberg Initiative Materials Science for Sustainability, Department of Environmental and Energy Sciences, Chalmers University of Technology, Chalmersplatsen 4, Gothenburg, SE-41296, Sweden
[2] Department of Environmental and Energy Sciences, Chalmers University of Technology, Chalmersplatsen 4, Gothenburg, SE-41296, Sweden
[3] Department of Physics and Astronomy, Chalmers University of Technology, Chalmersplatsen 4, Gothenburg, SE-41296, Sweden
*Corresponding author, stanicic@chalmers.se

# S1. Hydrogen production capacity

For chemical looping hydrogen production, it is desirable to have a metal oxide whose oxygen non-stoichiometry changes strongly between the reducing and oxidizing conditions, while being structurally stable and made of inexpensive, readily available elements. The number of oxygen vacancies per formula unit is denoted δ, and the non-stoichiometric compound is thus written as $ABO_{3-\delta}$. The δ-value is dependent on temperature and oxygen partial pressure, and it varies between the reactors where $\delta_{AR} \leq \delta_{SR} \leq \delta_{FR}$. In the three-step process, the following reactions take place in FR (1), SR (2) and AR (3).

$$ABO_{3-\delta_{AR}}(s) + (\delta_{FR} - \delta_{AR})\ CO\ (g) \rightarrow ABO_{3-\delta_{FR}}(s) + (\delta_{FR} - \delta_{AR})CO_2(g) \quad (1)$$

$$ABO_{3-\delta_{FR}}(s) + (\delta_{FR} - \delta_{SR})\ H_2O\ (g) \rightarrow ABO_{3-\delta_{SR}}(s) + (\delta_{FR} - \delta_{SR})H_2(g) \quad (2)$$

$$ABO_{3-\delta_{SR}}\ (s) + \frac{\delta_{SR} - \delta_{AR}}{2}\ O_2\ (g) \rightarrow ABO_{3-\delta_{AR}}(s) \quad (3)$$

Perovskite materials can release oxygen without undergoing phase transformation, which results in an oxygen deficiency in the material (δ). The free energy of the transition from $\delta_{FR}$ to $\delta_{SR}$ determines the hydrogen production capacity (HPC). Since the HPC is proportional to $\delta_{FR} - \delta_{SR}$, the free energy of this transition provides a thermodynamic descriptor of the material's hydrogen production potential. The free energy can be described as:

$$\Delta G_{\delta_{FR}-\delta_{SR}} = G_{f,ABO_{3-\delta_{SR}}} - G_{f,ABO_{3-\delta_{FR}}} - (\delta_{FR} - \delta_{SR})\ \mu_{O} \quad (4)$$

The chemical potential $\mu_O(T,p)$ can be derived from NIST-JANAF thermochemical tables[1], which can then be adjusted according to the specified reactor conditions. $\mu_O(T,p^0)$ is referenced to the DFT $O_2$ energy (see Table S1).

$$\mu_O(T,p) = \mu_O(T,p^0) + 0.5\ kTln\left[\frac{p_{O_2}}{p^0}\right] \quad 5$$

For reactions 1 and 2, the chemical potential of oxygen can be determined by the surrounding gas phase, i.e. the ratio $H_2O/H_2$ or $CO_2/CO$. In equilibrium with the gas phase ($H_2 + 0.5\ O_2 \rightleftharpoons H_2O$) the chemical potential can be described by $\mu_O = \mu_{H_2O} - \mu_{H_2}$. Hence, $\mu_O$ can be expressed in terms of $pH_2$, $pH_2O$ and equilibrium constant K(T) according to Eq. 6. The FR is treated the same way with $CO_2/CO$, using K(T) for $CO + 0.5\ O_2 \rightleftharpoons CO_2$ (from the JANAF values for $CO_2$ and CO). The AR uses $p_{O2}$ directly as in Eq. 5.

$$\mu_O(T,p) = \mu_O(T,p^0) + \ kT\left(\ln\left[\frac{p_{H_2O}}{p_{H_2}}\right] - lnK(T)\right) \quad 6$$

*Table S 1. Reactor conditions and corresponding oxygen chemical potentials used to evaluate the thermodynamic stability and oxygen non-stoichiometry of the perovskite materials. Each reactor, the fuel reactor (FR), steam reactor (SR) and air reactor (AR), is described by its temperature T and oxygen partial pressure $p_{O2}$ (relative to $p^o$ = 1 bar). The oxygen chemical potential is referenced to the DFT $O_2$ energy where $\mu_O = \Delta\mu_O + 0.5\ E(O_2)$, with $0.5\ E(O_2) = -5.961$ eV (DFT, Materials Project, 0 K).*

| | | FR | SR | AR |
|---|---|---|---|---|
| Temperature | °C | 900 | 800 | 1000 |
| Reactor condition | | $CO/CO_2 = 9$ | $H_2O/H_2 = 4$ | 5 vol. % $O_2$ in $N_2$ |
| $\Delta\mu_O(T,\mathrm{p})$ | eV | -3.415 | -3.018 | -1.615 |
| $\mu_O = \Delta\mu_O + \frac{1}{2}E(O_2)$ | eV | -9.375 | -8.979 | -7.576 |
| $p(O_2)$ | bar | $1\cdot10^{-18}$ | $7.1\cdot10^{-18}$ | $5\cdot10^{-2}$ |
| RT ln ($p(O_2)$) | kJ $mol^{-1}$ $O_2$ | -404 | -352 | -32 |

[1] T. C. Allison. NIST-JANAF Thermochemical Tables - SRD 13, doi: 10.18434/T42S31.

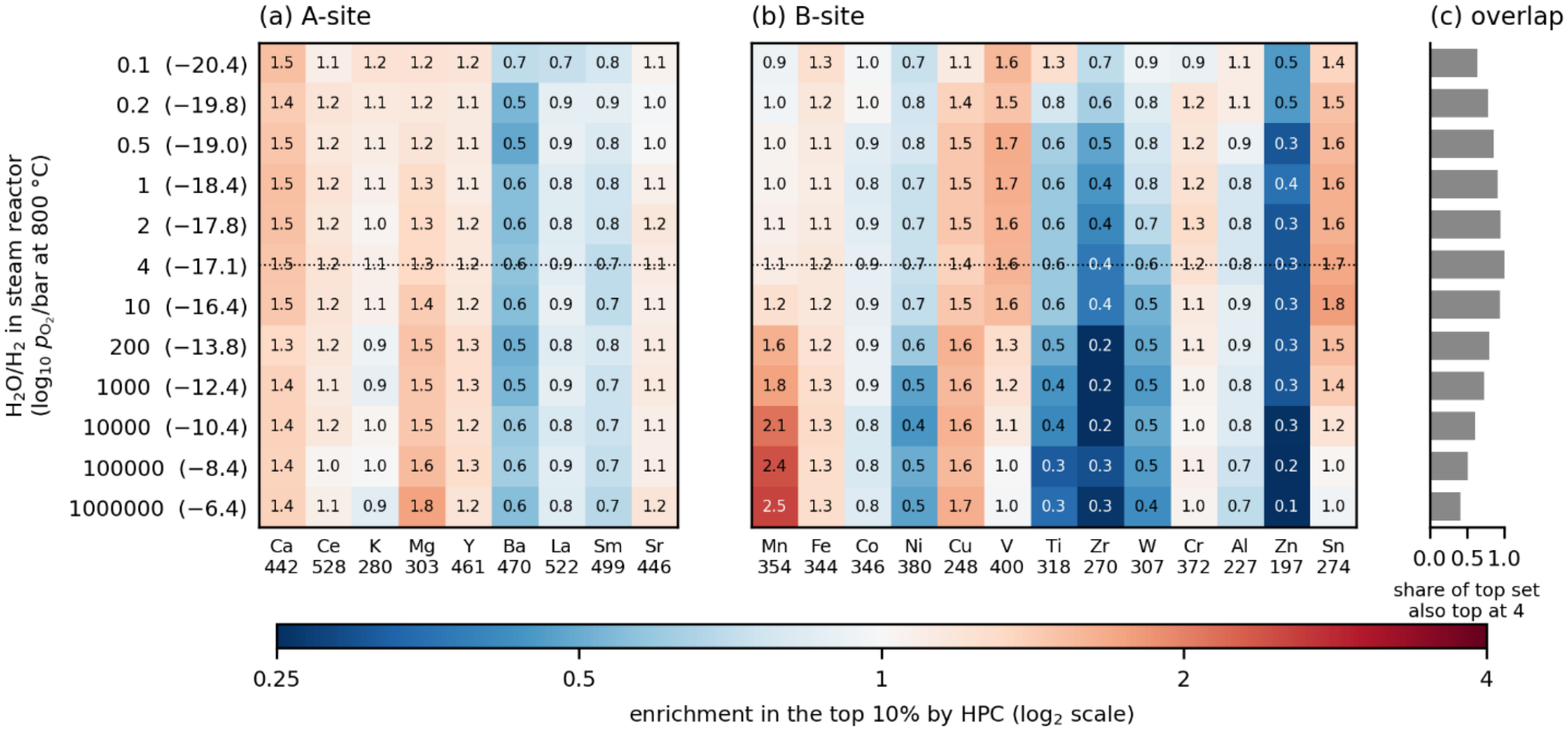

*Figure S 1. The HPC of the 2103 compounds of the starting set was recalculated for $H_2O/H_2$ ratios between 0.1 and $10^6$ in the steam reactor (800 °C), keeping the fuel reactor at the AL conditions ($CO_2/CO$ = 1/9, 900 °C). The corresponding equilibrium oxygen partial pressure of the steam reactor is given in brackets ($\log_{10}$ $pO_2$/bar). At each ratio the compounds were ranked by HPC and the top 10% (210 compounds) selected. Enrichment of each (a) A- and (b) B-site element in this top 10%. Colors are on a $\log_2$ scale: values above 1 (red) mean the element is favored at that ratio, values below 1 (blue) that it is disfavored. (c) Share of the top 10% at each ratio that is also in the top 10% at $H_2O/H_2$ = 4. The dotted line marks the conditions used during AL($H_2O/H_2$ = 4). $H_2O/H_2$ = $10^6$ corresponds to strongly oxidizing SR conditions, i.e. an equilibrium steam conversion of only ~0.0001%. With increasing $H_2O/H_2$ the top compounds shift towards Mn-rich B sites (enrichment rising from ~1.1 at $H_2O/H_2$ = 4 to ~2.5 at $10^6$).*

# S2. Active learning framework

Each active-learning cycle has four steps:

1. A deep ensemble of multilayer perceptrons (MLPs) is trained from scratch. It maps a cation composition to the Gibbs free-energy landscape over oxygen-vacancy content, for both the steam reactor (SR) and the fuel reactor (FR).
2. A Wasserstein autoencoder (WAE) is fine-tuned and used to generate candidate compositions.
3. The hydrogen production capacity (HPC) of each candidate is computed from each ensemble member's predicted landscape.
4. Ten candidates are chosen by Thompson sampling and passed on for SQS/phonon calculations.

Seven independent active-learning runs (r1–r7) were performed with random seeds 1–5, 42 and 100, each adding 10 compositions per cycle. Each composition is represented by its cation site fractions on the A sublattice (Ca, Ce, K, Mg, Y, Ba, La, Sm, Sr) and the B sublattice (Al, Cr, Co, Cu,

Fe, Mn, Ni, Sn, Ti, V, W, Zr, Zn). Each sublattice is normalized to sum to 1, giving a 22-dimensional input; oxygen is not included.

## S2.1. Training

Each member *m* is trained by randomized maximum-a-posteriori (MAP) sampling. This combines anchored ensemble[2] with the target perturbation of ensemble sampling[3] :

$$L_m = \left(\frac{1}{BD}\right) \Sigma_{i=1}^{B} \Sigma_{d=1}^{D} 0.5 \left[\log \sigma_{m,d}^2 + \frac{\left(\tilde{y}_{m,id} - \mu_{m,id}\right)^2}{\sigma_{m,id}^2}\right] + \left(\frac{1}{2\lambda N}\right) \left\|\theta_m - \tilde{\theta}_m\right\|^2$$

Here N is the training-set size, B the mini-batch size and D = 16 the number of outputs: G(v) in the SR and FR. The anchors $\tilde{\theta}_m$ are drawn once per member from $\mathcal{N}(0, \lambda I)$ with λ = 1 for all weight tensors, including the LayerNorm gains. The bias anchors are zero, so the biases are regularized toward zero. The perturbed targets are $\tilde{y}_m = y + \epsilon_m$, where $\epsilon_m$ is drawn once per member and training sample and kept fixed. Its standard deviation rises linearly with vacancy count, from 0.2 eV at v = 0 to 0.8 eV at v = 8, and is the same for SR and FR. The targets are normalized with one standard deviation shared by all G(v) (7.331 eV), so the noise is 0.027–0.109 in normalized units. The predicted log-variance $\log \sigma^2$ is clamped to [−5, 1] in normalized units. All members are trained on the full training set without bootstrapping. Diversity between members comes from the anchors, the target perturbations, the random initialization and the independent mini-batch shuffling. The training hyperparameters are given in Table S 2.

*Table S 2. Ensemble architecture and training hyperparameters.*

| **Hyperparameter** | **Value** |
|---|---|
| Ensemble size | 16 members |
| Layers | 22 → 64 → 64, each Linear → LayerNorm → SiLU → Dropout(0.25) |
| Output heads | Two linear heads (64 → 16): a mean μ and a log-variance $\log \sigma^2$ for each target |
| Parameters | 7,968 per member, 127,488 in total |
| Initialization | PyTorch default, with a different seed for each member |
| Optimizer | Adam, learning rate $1\times10^{-4}$, no weight decay |
| Batch size | 64 (independent shuffling for each member) |
| Maximum epochs | 500 |
| LR schedule | ReduceLROnPlateau for each member on its own validation NLL (factor 0.5, patience 50) |

[2] Pearce, T., et al. (2020). Uncertainty in neural networks: Approximately Bayesian ensembling. International conference on artificial intelligence and statistics, PMLR.
[3] Lu, X. and B. Van Roy (2017). "Ensemble sampling." Advances in neural information processing systems **30**.

| | |
|---|---|
| Gradient clipping | Global norm 1.0 for each member |
| Validation split | 15 % random |
| Early stopping | On the mean validation NLL across members, patience 150; best weights restored |
| Retraining | From scratch in every cycle |

## S2.2. Generative model (WAE)

A Wasserstein autoencoder (WAE) maps each composition and its z-scored HPC to an 8-dimensional latent space and decodes it back to site fractions and HPC. It is trained in cycle 1, fine-tuned in each later cycle, and used to generate new candidate compositions (Section S2.3). The architecture and training settings are given in Table S 3.

*Table S 3. WAE architecture and training.*

| **Item** | **Value** |
|---|---|
| Input (23D) | z-scored HPC + 22 site fractions |
| Encoder | 23 → 320 → 256 → 192 → 8 (latent), each hidden layer Linear → LayerNorm → SiLU → Dropout(0.1) |
| Composition decoder | 8 → 192 → 256 → 320 → 22, same layer type, separate softmax over each sublattice |
| Property decoder | 8 → 96 → 128 → 160 → 1 (HPC), same layer type |
| Parameters | 319,103 |
| Loss | $L_1$ reconstruction of all 23 outputs + + $\lambda_{MMD} \cdot \mathrm{MMD}(q(z), \mathcal{N}(0, I))$ /B |
| MMD kernel | Inverse multiquadratic, $k = C/(C + \lVert z - z' \rVert^2)$, $C = 2 \cdot d \cdot s$, $d = 8$, $s \in \{0.1, 0.2, 0.5, 1, 2, 5, 10\}$; $\lambda_{MMD} = 0.01$ |
| Optimizer | Adam, weight decay $1\times10^{-5}$, batch size B=32, gradient clipping 1.0 |
| Cycle 1 | 1500 epochs, learning rate $1\times10^{-3}$, ReduceLROnPlateau on validation loss (factor 0.5, patience 100) |
| Later cycles | Warm-started, 500 epochs, learning rate $2\times10^{-4}$, scheduler patience 50 |
| Early stopping | Patience 800 on validation loss |

## S2.3. Candidate generation, HPC and sampling

To generate candidates, a single full-covariance Gaussian is fitted to the WAE latent vectors of the training set, and 25,000 points are drawn from it and decoded into compositions. Each composition is rounded to whole atom counts (16 per sublattice). To find the optimal vacancy content, dropout is turned off and, for each ensemble member, the predicted mean landscape is denormalized and a quadratic $G(v) = av^2 + bv + c$ is fitted over v = 0–8, separately for SR and FR. The optimum $v^*$ is the vertex clipped to [0, 7]. If the parabola opens downward ($a < 0$), the endpoint (v = 0 or 8) with the lower energy is used instead; if the fit is essentially linear ($|a| < 10^{-10}$), the endpoint is chosen by the sign of the slope.

Each member's HPC is then computed as $HPC = 1000\,\frac{\delta_{FR}-\delta_{SR}}{M_{ABO_3}}$ where $M_{ABO_3}$ is the molar mass of the supercell composition in g mol$^{-1}$, giving HPC in mmol g$^{-1}$. The 16 member values form the predictive distribution, and its mean and standard deviation are reported. Candidates are selected by Thompson sampling: for each candidate, one ensemble member is picked at random and its HPC is used as the score. Candidates are de-duplicated at 1/80 compositional resolution, and the 10 with the highest scores are selected.

## S2.4. Visualization

To follow how the selected compositions are distributed in the learned representation, we projected the latent space to two dimensions with Uniform Manifold Approximation and Projection (UMAP)[4]. To relate the regions of the map to chemistry, the same coordinates were colored by HPC and by the dominant A- and B-site elements (Figure S 2). Compositions in which two or more elements share the largest count are marked as ties. To keep colors distinguishable, only eight of the most frequent dominant elements among the AL-selected compositions are shown individually on each sublattice. The rest are groups as "other". The map separates into regions of distinct chemistry, and the high HPC-compositions concentrate in a Sn-dominated region with mainly K and Mg on the A-site.

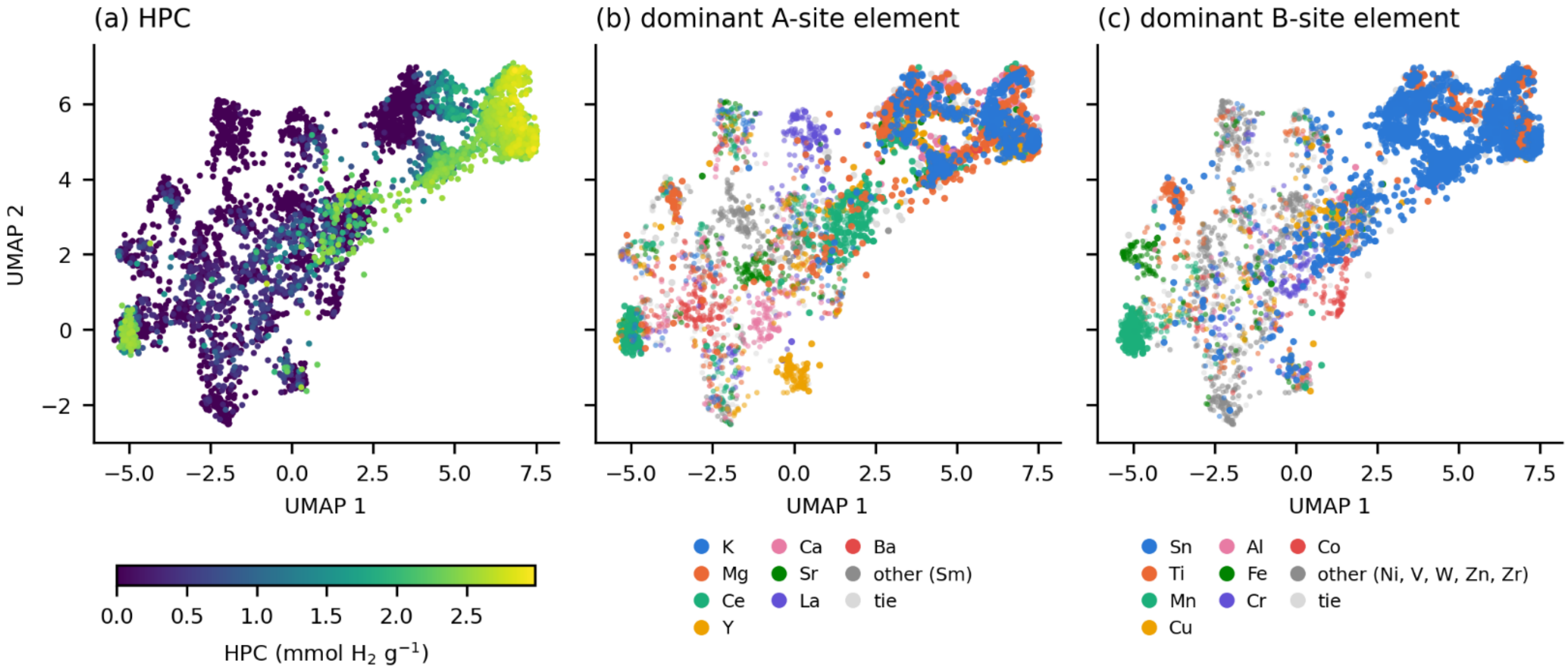

*Figure S 2. Two-dimensional UMAP projection of the 8-dimensional latent representations learned by the WAE, on the same map as Figure 4. Each point represents a candidate perovskite composition embedded according to its learned latent representation. Points lie close together when their latent representations are similar, so distinct regions reflect groups of similar compositions. (a) Colored by HPC, (b) colored by the dominant A-site element and (c) by the dominant B-site element, i.e., the element occupying the most sites of that sublattice. Points where two or more elements share the largest count are shown in light grey ("tie"). Elements that are rarely dominant are pooled as "other": Sm on the A-site; Ni, V, W, Zn and Zr on the B-site. In (b) and (c), starting-set compositions are drawn smaller and lighter than those selected by active learning.*

---

[4] McInnes, L., et al. (2018). "Umap: Uniform manifold approximation and projection for dimension reduction." arXiv preprint arXiv:1802.03426.

## S2.5. Linking composition to redox capacity

For each composition, the vacancy-resolved Gibbs energies $G_v(T)$ of the 80-atom SQS cell (v = 0–8, δ = v/16) were used. At each temperature, $G_v$ was fitted with a quadratic function, $G_v \approx a v^2 + b v + c$. The equilibrium vacancy count at oxygen chemical potential $\mu_O$ is the minimum of $G_v + v\,\mu_O$, i.e. $v^* = -(b + \mu_O)/(2a)$. The reduction midpoint is the $\mu_O$ at which $v^* = 4$ (δ = 0.25), $\mu_O = -(8a + b)$. It is reported as $RT \ln pO_2 = 2[\mu_O - \mu_O°(T)]$, in kJ mol$^{-1}$ $O_2$, for 600–1200 °C. All charge-neutral AL compositions (starting set excluded) with τ ≤ 4.18 were considered. The midpoint y at 800 and 900 °C was fitted by least squares (no intercept) against the per-cell element counts $y = \sum_A \beta_a n_a + \sum_B \gamma_b n_b$ where a runs over Ca, Sr, Ba, Mg, K, La, Y, Sm, Ce, and b runs over Al, Cr, Co, Cu, Fe, Mn, Ni, Sn, Ti, V, W, Zr, Zn. Since $\sum_A n_a = \sum_B n_b = 16$ for every composition, individual coefficients are not unique, and we report only the within-site contrasts:

$$\Delta_a = \beta_a - \bar{\beta}_A, \text{ with } \bar{\beta}_A = \Sigma_{\{i \in A\}} w_i \beta_i \text{ and } w_i = \frac{\langle n_i \rangle}{16},$$

Where $\langle n_i \rangle$ is the mean number of i atoms per 16 A sites over the 2712 compositions. $\Delta_b$ being defined analogously for the B-site fractions. $\Delta_a$ is the change in midpoint (kJ mol$^{-1}$ $O_2$) when an average A-site cation is replaced by element a. On the A site, the mean occupancy is about 3.5 K, 3.1 Mg, 3.0 Ce, 2.6 Y and 1.6 Ca per 16 sites, with the remaining ~2.3 shared among Sr, La, Sm and Ba. On the B site it is about 8.4 Sn, 3.5 Ti and 1.2 Al per 16 sites, with the remaining ~2.9 shared among Mn, Cu, V and minor amounts of other transition metals, Zr and Zn. A contrast of zero therefore means no change relative to this dataset average. Differences between contrasts on the same site, and hence the charge-neutral substitutions discussed in the main text, do not depend on this choice of reference.

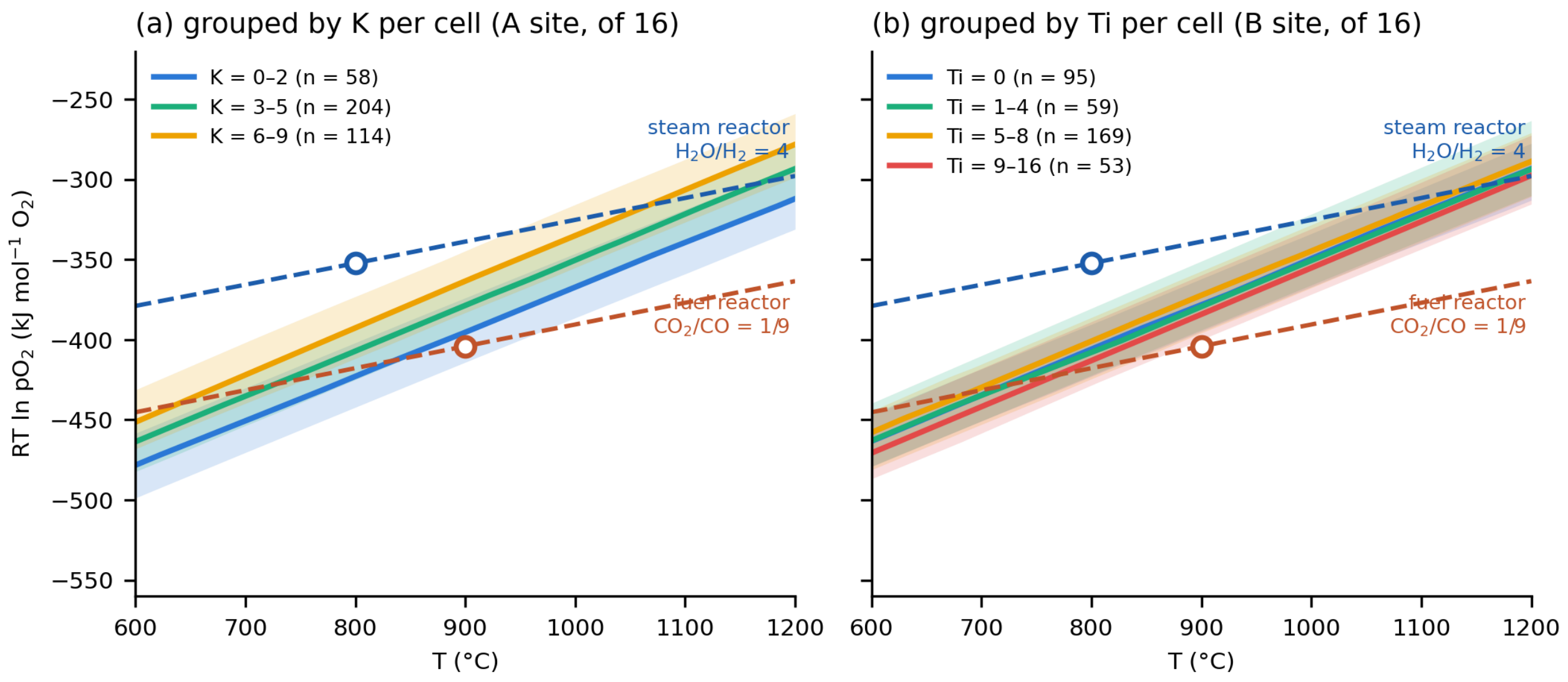

*Figure S 3. Ellingham diagram of the reduction midpoint (δ = 0.25) of 376 Sn-rich AL compositions (≥ 6 Sn per 80-atom cell), grouped by (a) K and (b) Ti content: median (line) and 25–75 % range (band). The reduction midpoint of a composition is the oxygen chemical potential ($RT \ln pO_2$) at which its equilibrium oxygen non-stoichiometry is δ = 0.25, halfway between the fully oxidized state (δ = 0) and the largest non-stoichiometry considered (δ = 0.5). Above its line a composition is predominantly oxidized, below it predominantly reduced. Dashed lines: steam reactor ($H_2O/H_2$ = 4) and fuel reactor ($CO_2/CO$ = 1/9); circles: operating*

*temperatures. With increasing K, the median midpoint at 800 °C rises from −423 kJ mol $^{-1}$ $O_2$ (0–2 of 16 A-sites) to −393 kJ mol $^{-1}$ $O_2$ (6–9 of 16 A-sites), into the window between the two reactors, whereas with Ti it stays between −401 and −413 kJ mol $^{-1}$ $O_2$ for 0 to 16 Ti per cell.*

# S3. Multi-objective ranking strategy

Candidate materials were ranked by predicted performance and by practical considerations relevant to large-scale implementation. Four metrics were considered: hydrogen production capacity (HPC), raw material price, supply resilience, and the potential environmental and health index of the constituent oxides. Because the metrics are on different numerical scales, each was converted to a percentile rank. The metrics weighed 0.6 (HPC), 0.1 (price), 0.15 (environmental and health impact) and 0.15 (supply resilience). This prioritizes performance while still accounting for scalability, environmental considerations and cost.

## S3.1. Environmental and health (EH) index

To discourage compositions with hazardous oxide constituents, an element-level EH hazard index was developed. It ranges from 1 (no CLP hazard classification) to 0 (most severe hazard profile). For each cation, binary oxides representing its common oxidation states were considered, and the element was assigned the index of its most hazardous oxide. For chromium, Cr(VI) oxide was included as a conservative case, because hexavalent chromium is highly hazardous and chromates can form when Cr is oxidized together with alkaline-earth or alkali oxides at high temperature. The index is a comparative screening metric for material selection, not a regulatory classification or a quantitative measure of toxicity.

Hazard data for each oxide was taken from the EU CLP Regulation (EC) No 1272/2008. Where a harmonized classification exists in Annex VI, Table 3 (consolidated up to the 23rd Adaptation to Technical Progress, Commission Delegated Regulation (EU) 2025/1222)[5], it was used, since it is legally binding and based on peer review by ECHA's Risk Assessment Committee. This applied to ZnO, $MnO_2$, $Cu_2O$, CuO, CoO, NiO, $Ni_2O_3$, $V_2O_5$ and $CrO_3$. For oxides without a harmonized entry, suppliers are required to self-classify (CLP Art. 4) and notify the classification to the ECHA Classification and Labelling (C&L) Inventory (CLP Art. 40)[6]. Because notified classifications for the same substance often differ between notifiers, a hazard statement was included only if at least 50 % of the notifications reported it (base case). A precautionary scenario with a 25 % threshold was also evaluated; to assess how sensitive the ranking is to this choice. Aggregated notification data were retrieved from PubChem[7], which compiles the ECHA C&L Inventory (accessed 6 October 2026).

---

[5] European Chemicals Agency (ECHA). Table of harmonized entries in Annex VI to CLP (consolidated up to ATP23). https://echa.europa.eu/information-on-chemicals/annex-vi-to-clp (accessed 6 October 2026).
[6] European Chemicals Agency (ECHA). Classification and Labelling (C&L) Inventory. https://echa.europa.eu/information-on-chemicals/cl-inventory-database (accessed 6 October 2026).
[7] National Center for Biotechnology Information. PubChem Compound Database, GHS Classification section (aggregated from ECHA C&L notifications). https://pubchem.ncbi.nlm.nih.gov (accessed 6 October 2026).

Each hazard statement was assigned a severity level from 1 to 5 within four domains including chronic health (C), acute/local health (A), environment (E) and physical hazards (P) (Table S 4).

*Table S 4. Severity levels assigned to CLP hazard classes and categories (with hazard statement codes) in four hazard domains: chronic health (C), acute/local health (A), environment (E) and physical hazards (P). Level 5 is the most severe and is reserved for carcinogenicity, mutagenicity and reproductive toxicity in categories 1A/1B and for respiratory sensitization, following the criteria for substances of very high concern in REACH Art. 57. The M-factor is the multiplying factor assigned to highly toxic substances for aquatic toxicity. The highest level in each domain is used to calculate the hazard score HS (Section S3.1).*

| Level | Chronic health (C) | Acute/local health (A) | Environment (E) | Physical (P) |
|---|---|---|---|---|
| 5 | Carc./Muta./Repr. 1A/1B (H350, H340, H360); Resp. Sens. 1 (H334) | – | – | – |
| 4 | – | Acute Tox. 1–2 (H300/310/330) | Aquatic 1 with M ≥ 10 | – |
| 3 | STOT RE 1 (H372); Carc./Muta./Repr. 2 (H351/341/361) | Acute Tox. 3 (H301/311/331); Skin Corr. 1 (H314); Eye Dam. 1 (H318) | Aquatic Acute/Chronic 1 (H400/H410) | Ox. Sol. 1 (H271) |
| 2 | STOT RE 2 (H373); Skin Sens. 1 (H317); Lact. (H362) | Acute Tox. 4 (H302/312/332) | Aquatic Chronic 2 (H411) | Ox. Sol. 2–3 (H272) |
| 1 | – | Skin/Eye Irrit. 2; STOT SE 3 (H315/319/335) | Aquatic Chronic 3–4 (H412/413) | Met. Corr. 1 (H290) |

Physical hazards were included because they are relevant to handling at scale. The highest level was reserved for carcinogenicity, mutagenicity and reproductive toxicity in categories 1A/1B and for respiratory sensitization, consistent with the criteria for substances of very high concern in REACH Art. 57[8]. For aquatic toxicity, the M-factor (a multiplying factor assigned to highly toxic substances) was used to distinguish very high from high environmental hazard. For each oxide, a hazard score was calculated as $HS = L_{max} + w\sum L_{other}$ with $w$ =0.25, and where $L_{max}$is the highest domain level, and $\sum L_{other}$ is the sum of the highest levels in the remaining domains. The worst hazard thus dominates, while additional hazards in other domains lower the score. The score was normalized to an EH index, $EH = 1 - \frac{HS}{HS_{max}}$ where $HS_{max} = 7.75$ is the theoretical maximum. Hence, EH = 1 means no classification and EH = 0 the most severe element (Table S 5). The ranking was recalculated for $w$ = 0 (worst hazard only) and $w$ = 0.5, and for the 25 % notification threshold. The grouping into hazard tiers (Table S 5) was preserved in all cases; only the order within tiers changed. The approach reflects intrinsic hazard only. Exposure, particle size, solubility and bioavailability are not considered, and self-classifications carry greater uncertainty than harmonized classifications. No classification data were available for $Co_3O_4$, so the index for Co is based on CoO.

8

*Table S 5. Environmental and health (EH) hazard index of the cations, based on the most hazardous of their binary oxides. EH ranges from 1 (no CLP classification) to 0 (most severe theoretical hazard profile). Classification data: harmonized classifications from Annex VI of the CLP Regulation (ATP23) were used where available. Otherwise, hazards notified to the ECHA C&L Inventory were used, included at ≥ 50 % of notifications (base case) or ≥ 25 % (precautionary case). Percentages in parentheses give the share of notifications reporting the hazard. Tiers group elements with similar index values and are unchanged across the sensitivity cases (w = 0, 0.25 and 0.5; 25 % and 50 % thresholds). "Uncertain" marks elements that are unclassified in the base case but have hazards reported by a substantial minority of notifiers. Cr is represented by $CrO_3$ as a conservative case for Cr(VI) formation. Co is based on CoO only, because no classification data is available for $Co_3O_4$.*

| Tier | Element | EH (base) | EH (precautionary) | Driving oxide and hazards | Data basis |
|---|---|---|---|---|---|
| Low | Mg, Al, Zr | 1.00 | 1.00 | none | notifications (>90 % not classified) |
| | Fe | | 0.87 | $Fe_3O_4$: H315, H319, H335 (~32 %) | notifications |
| | La, Y | | 0.87 | H315, H319, H335 (~40 %) | notifications |
| Uncertain | Ti | | 0.61 | $TiO_2$: H351 (36 %) | notifications (harmonized entry annulled) |
| | Sm | | 0.61 | $Sm_2O_3$: H400, H410 (45 %) | notifications |
| | Sn | | 0.55 | SnO: H372, H302, H317 (35 %) | notifications |
| Moderate | W | 0.74 | 0.74 | $WO_3$: H302, H315, H319, H335 (57 %) | notifications |
| | Mn | 0.74 | 0.61 | $MnO_2$: Acute Tox. 4; $Mn_3O_4$: H361 (49 %) | harmonized + notifications |
| | Ca, Sr, K | 0.61 | 0.61 | Eye Dam. 1 / Skin Corr. 1 | notifications (78–93 %) |
| | Zn | 0.61 | 0.61 | ZnO: H400, H410 | harmonized |
| | Ce | 0.61 | 0.61 | $Ce_2O_3$: H400, H410 (91 %); $CeO_2$ not classified | notifications |
| | Ba | 0.61 | 0.52 | BaO: H314 (51 %); H271 (37 %) | notifications |
| High | Cu | 0.39 | 0.39 | $Cu_2O$: Acute Tox. 4, Eye Dam. 1, Aquatic 1 (M = 100) | harmonised |
| | Co | 0.36 | 0.36 | CoO: Acute Tox. 4, Skin Sens. 1, Aquatic 1 (M = 10) | harmonized (no data for $Co_3O_4$) |
| | Ni | 0.32 | 0.32 | NiO: Carc. 1A, STOT RE 1, Skin Sens. 1 | harmonized |
| Very high | V | 0.16 | 0.16 | $V_2O_5$: Carc. 1B, Acute Tox. 2, STOT RE 1 | harmonized |
| | Cr | 0.03 | 0.03 | $CrO_3$: Carc. 1A, Muta. 1B, Acute Tox. 2, Ox. Sol. 1 | harmonized |

## S3.2. Supply resilience index

To account for raw material availability, each cation was assigned a supply resilience score based on the supply risk (R) indicator of the fifth EU criticality assessment (2023), which underlies the list of critical raw materials in the Critical Raw Materials Act (Regulation (EU) 2024/1252)[9]. R represents supply risk from an EU perspective. For each element, the R of the raw material most relevant to oxide precursor production was used, at the stage used for R in the assessment: for example, magnesite rather than magnesium metal for Mg, and titanium ores rather than titanium metal for Ti. For Co, V, Ce and Y, EU sourcing data were not available in the assessment, and R is based on global supply only. For Sr, the EU has no import reliance, but EU sourcing depends almost entirely on a single country (Spain, 99 %) with no end-of-life recycling, which results in a comparatively high R despite significant domestic production. The supply resilience score was calculated as $Res = 1 - R/R_{max}$, where $R_{max} = 4.0$ is the highest R among the elements considered (Ce), so that 1 corresponds to the lowest and 0 to the highest supply risk (Table S 6).

[9] Regulation (EU) 2024/1252 of the European Parliament and of the Council of 11 April 2024 establishing a framework for ensuring a secure and sustainable supply of critical raw materials [2024] OJ L1252.

*Table S 6. Supply resilience score of the cations, derived from the supply risk (R) indicator of the EU 2023[10] criticality assessment. Res = 1 − R/Rmax (Rmax = 4.0, Ce). For each element, the raw material most relevant to oxide precursor production was used, at the stage (extraction or processing) used for R in the assessment. Supply basis: GS+EU = global supply and EU sourcing; EU = EU sourcing only; GS = global supply only. CRM = critical raw material; SRM = strategic raw material. Main global supplier and EU import reliance are taken from the annexes of the assessment; the supplier is given for CRMs and SRMs only.*

| Element | Raw material | Stage | Supply basis | R | EU SRM/CRM 2023 | Main global supplier | Import Reliance | Supply resilience score (Res) |
|---|---|---|---|---|---|---|---|---|
| Cu | Copper | Extraction | GS+EU | 0.1 | SRM | Chile 28 % | 48% | 0.98 |
| Zn | Zinc | Extraction | GS+EU | 0.2 | – | – | | 0.95 |
| Ca | Limestone | Extraction[2] | EU | 0.3 | – | – | | 0.93 |
| Fe | Iron ore | Extraction | GS+EU | 0.5 | – | – | | 0.88 |
| Ti | Titanium[11] | Extraction | GS+EU | | – | – | | |
| Ni | Nickel | Processing | GS+EU | | SRM | China 33 % | 75% | |
| Mg | Magnesite[12] | Extraction[2] | EU | 0.6 | – | – | | 0.85 |
| K | Potash | Extraction | GS+EU | 0.7 | – | – | | 0.83 |
| Cr | Chromium | Extraction | GS | | – | – | | |
| Zr | Zirconium | Extraction | GS+EU | 0.8 | – | – | | 0.80 |
| Sn | Tin | Processing | GS+EU | 0.9 | – | – | | 0.78 |
| Al | Aluminium/bauxite | Extraction | GS+EU | 1.2 | CRM | Australia 28 % | 89% | 0.70 |
| Mn | Manganese | Extraction | GS+EU | | CRM | S. Africa 29 % | 96% | |
| W | Tungsten | Processing | GS+EU | | CRM | China 86 % | 80% | |
| Ba | Baryte | Extraction | GS+EU | 1.3 | CRM | China 32 % | 74% | 0.68 |
| V | Vanadium | Extraction | GS | 2.3 | CRM | China 62 % | No production nor trade of V-ores and concentrates in EU | 0.43 |
| Sr | Strontium | Extraction | GS+EU | 2.6 | CRM | Iran 37 % | 0% | 0.35 |
| Co | Cobalt | Extraction | GS | 2.8 | CRM | DRC 63 % | 81% | 0.30 |
| La | Lanthanum (LREE) | Processing | GS+EU | 3.5 | CRM | China 85 % | 100% | 0.13 |
| Sm | Samarium (LREE) | Processing | GS+EU | | CRM | China 85 % | 100% | |
| Y | Yttrium (HREE) | Processing | GS | | CRM | China 100 % | 100% | |
| Ce | Cerium (LREE) | Processing | GS | 4.0 | CRM | China 85 % | 100% | 0.00 |

[10] European Commission: Directorate-General for Internal Market, Industry, Entrepreneurship and SMEs, Grohol, M. and Veeh, C., Study on the critical raw materials for the EU 2023 – Final report, Publications Office of the European Union, 2023, https://data.europa.eu/doi/10.2873/725585

[11] Titanium" refers to ilmenite and rutile, the feedstock for $TiO_2$. Titanium metal is a separate CRM with SR = 1.6.

[12] MgO is made by calcining magnesite. Magnesium metal is a CRM with SR = 4.1 and China supplying 91 %; using it would drop Mg to 0.